\documentclass[twocolumn,english,aps,superscriptaddress,showpacs,pre,floatfix,nofootinbib,longbibliography]{revtex4-1}
\usepackage{mathptmx}

\usepackage[latin9]{inputenc}
\usepackage{color}
\usepackage{babel}
\usepackage{amsmath}
\usepackage{amssymb}
\usepackage{graphicx}
\usepackage{esint}
\usepackage[unicode=true,pdfusetitle,
 bookmarks=true,bookmarksnumbered=false,bookmarksopen=false,
 breaklinks=true,pdfborder={0 0 0},pdfborderstyle={},backref=false,colorlinks=true]
 {hyperref}
\hypersetup{
 linkcolor=blue,citecolor=blue,urlcolor=blue}

\makeatletter

\providecommand{\tabularnewline}{\\}

\usepackage{babel}

\makeatother

\begin{document}
\title{Effect of long-range {hopping} on dynamic quantum phase
transitions of an exactly solvable free-fermion model: non-analyticities
at almost all times}
\author{J. C. Xavier}
\affiliation{Universidade Federal de Uberlândia, Instituto de Física, C.~P.~593,
38400-902 Uberlândia, MG, Brazil}
\author{José A. Hoyos}
\affiliation{Instituto de Física de São Carlos, Universidade de São Paulo, C.~P.~369,
São Carlos, São Paulo 13560-970, Brazil}
\affiliation{Max Planck Institute for the Physics of Complex Systems, N\"othnitzer
Str. 38, 01187 Dresden, Germany}
\begin{abstract}
In this work, we investigate quenches in a free-fermion chain with
long-range hopping which decay with the distance with an exponent
$\nu$ and has range $D$. By exploring the exact solution of the
model, we found that the dynamic free energy is non-analytical, in
the thermodynamic limit, whenever the sudden quench crosses the equilibrium
quantum critical point. We were able to determine the non-analyticities
of dynamic free energy $f(t)$ at some critical times $t^{c}$ by
solving nonlinear equations. We also show that the Yang-Lee-Fisher
(YLF) zeros cross the real-time axis at those critical times. We found
that the number of nontrivial critical times, $N_{s},$ depends on
$\nu$ and $D$. In particular, we show that for small $\nu$ and
large $D$ the dynamic free energy presents non-analyticities in any
time interval $\Delta t\sim1/D\ll1$, i.e., there are \emph{non-analyticities
at almost all times}. For the spacial case $\nu=0$, we obtain the
critical times in terms of a simple expression of the model parameters
and also show that $f(t)$ is non-analytical even for finite system
under anti-periodic boundary condition, when we consider some special
values of quench parameters. We also show that, generically, the first
derivative of the dynamic free energy is discontinuous at the critical
time instant when the YLF zeros are non-degenerate. On the other hand,
when they become degenerate, all derivatives of $f(t)$ exist at the
associated critical instant. 
\end{abstract}
\date{\today}
\maketitle

\section{Introduction}

Equilibrium phase transitions (PTs) have been detail studied and
observed in several compounds in the last two centuries~\citep{booksachdev,bookStanley,RevPT1997}.
Along the lines (or planes, or points) that separate the distinct
phases, the thermodynamic functions are non-analytic. Due to this
fact, the systems present unusual physical properties close to these
lines. In general, we can not understand the phenomena close to the
transition lines by a simple picture, such as the Fermi liquid for
instance. For this reason, this subject has been of great interest
for several physicist communities. The non-analyticity of the thermodynamic
functions is encoded in the zeros of the partition function ${\cal Z}$,
the so-called Yang-Lee-Fisher (YLF) zeros~\citep{YLzerosI,YLzerosII,bookFisherYLzeros}.
In general, the zeros of ${\cal Z}(q)=\text{Tr}\left(e^{-qH}\right)$
happen for $q=\beta+i\alpha$, where $\beta=\frac{1}{k_{B}T}$ and
$\alpha\neq0$. In the thermodynamic limit, however, these zeros can
touch the real temperature axis yielding to non-analyticities of the
Helmholtz free energy $F=-k_{B}T\ln\left({\cal Z}(\beta)\right)$.
For a recent experimental verification of this phenomenon, see, for
instance, Refs.~\citealp{PRLYLFzeros2012,PRLYLFzeros2015}. A similar
equilibrium partition function that is also studied is the boundary
partition function $\mathcal{Z}^{b}(\beta)=\left\langle \psi^{b}\left|e^{-\beta H}\right|\psi^{b}\right\rangle $,
i.e., the partition function ruled by the Hamiltonian $H$ with boundaries
described by the boundary state $\left|\psi^{b}\right\rangle $ separated
by $\beta$~\citep{Cardyboundary,cardyboundary2,LECLAIboundary}. 

In the last years~\citep{HeylPRLseminalDynamic,DynamSirkerPRB2014,DynamicVajnaPRB2014,KarraschPhysRevB.87.195104,FirtOderDynanimcPRL,PRBVajnaDynamic,DynamicLongRangePRB2017,PRLHeylLongRangDynamic,DQPTexpnatur2018,dynamicDelgadoPhysRevB.99.054302,SciRepJafari,DQPTexpPhysRevApplied.11.044080,DQPTlongrangePhysRevE.96.062118,heyl-trappexp,DQPTlograngisingPhysRevB.96.104436,NetoRafaelXavierPRBL2022},
the concept of YLF zeros has been applied to sudden quenches: a parameter
$\delta$ of a system Hamiltonian $H\left(\delta\right)$ changes
from $\delta_{0}\rightarrow\delta$ at the time instant $t=0$. Specifically,
the dynamical analog of the boundary partition function is the return
probability $Z(t)=\left\langle \psi_{0}\left|e^{-iH(\delta)t}\right|\psi_{0}\right\rangle $,
where $\left|\psi_{0}\right\rangle $ is the ground state of the Hamiltonian
$H\left(\delta_{0}\right)$. The dynamical analog of the free energy
is $f(t)=-\frac{1}{N}\ln\left(\left|Z(t)\right|^{2}\right)$, where
$N$ is the number of degrees of freedom, and can also be a non-analytic
function at some critical time $t^{c}$. For a review and generalizations
to other out-of-equilibrium scenarios see, e.g., Ref.~\citealp{dynampt}.

The quantum quench protocol we consider here is the following: the
system is prepared in the ground state of $H(\delta_{0})$ and then
is time-evolved according to $H(\delta)$, being $\delta$ some tuning
parameter of $H$. The non-analytical behavior of $f$ in time was
called dynamical quantum phase transition (DQPT)~\citep{HeylPRLseminalDynamic}
and was recently observed in experiments~\citep{heyl-trappexp,DQPTexpPhysRevApplied.11.044080,DQPTexpnatur2018}.
It is important to mention that, by now, it is well established that
there is no one-to-one correspondence between DQPTs and equilibrium
phase transitions~\citep{DynamSirkerPRB2014,DynamicVajnaPRB2014,FirtOderDynanimcPRL,PRBVajnaDynamic,DynamicLongRangePRB2017,PRLHeylLongRangDynamic,SciRepJafari}. 

{Experimental observation of the DQPTs was observed
recentely ~\citep{heyl-trappexp}, where trapped ions were used to
simulate the transverse-field Ising chain with long range interaction.
The long range interaction between two spins $i$ and $j$ is given
by $J_{i,j}=\Omega^{2}\nu_{R}\sum_{m}\frac{b_{im}b_{jm}}{\mu^{2}-\nu_{m}^{2}}$
\citep{IslamTrapIons2013} and depends on the experimental setup,
namely: the Rabi frequency $\Omega$ of the laser, the ion mass of
the single ion via the recoil frequency $\nu_{R}$ associated with
the dipole force, the orthonormal mode component of the ith ion $b_{im}$
with mode $m$ and frequency $\nu_{m}$, as well as the symmetric
detuning $\mu$ of the beatnote from the spin-flip transition ~\citep{PorrasCiracTrapIons,IslamTrapIons2013,heyl-trappexp,RMPTrap2021,Marino_2022Review}.
It has been observed in trapped ion experiments that the long range
coupling $J_{i,j}$ can be approximated as $J_{i,j}\sim1/|i-j|^{\alpha}$
where $0<\alpha<3$ depends on the laser detuning $\mu$ \citep{IslamTrapIons2013,heyl-trappexp,ZhangTrapIons2017,JoshiTrapIons2022}.}

The effect of the long-range interactions in the context of the DQPTs
were investigated in the transversal-Field Ising chain~\citep{heyl-trappexp,DQPTlograngisingPhysRevB.96.104436,DynamicLongRangePRB2017,DQPTlongrangePhysRevE.96.062118,PRLHeylLongRangDynamic}.
All those studies were done numerically since the long-range interaction,{
in general, breaks integrability (exceptions exist and can be found
in, e.g., \onlinecite{Bojan-LMGmodel,Kosier-LMGmodel}).} Although
numerical results can give strong evidence of the DQPTs, those methods
are limited. In particular, the studies based on exact diagonalization
and/or matrix product state (MPS) are limited by the size of the system,
and/or by the bond dimension, as well as limited to short times. In
principle and strictly according to the YLF zeros theory, the DQPTs
manifest only in the thermodynamic limit. In this sense, a rigorous
proof of the existence of a DQPT in the transversal-Field Ising chain
with long-range interaction is still missing. In this vein, it is
highly desirable to have a deep understanding of the long-range interaction
effects in the context of DQPTs through analytical results. {Insights
into this issue may be gained by considering the free fermions with
long-range hopping, since the model can be mapped, by using the Jordan-Wigner
transformation, in a XX chain with long range interaction. Although
in this case, multiple spin interactions appear }\citep{suzukixy,Dalsonxavier,LongRangeJones}.
{Very recently, the effect of the long range hopping
in the context of the DQPT were investigated in few models, like 
some variant of the Kitaev chain \citep{KitaevLongRangQDPT2017,KitaevLongRangeHalimeh,KitaevLongHalimedsupercon}
(see also Refs. {\onlinecite{Bojan-LMGmodel,Kosier-LMGmodel}}).}
Motivated by the aforementioned facts, we investigate DQPTs in an
exactly solvable free fermion model with long-range hoppings.

The paper is organized as follows: In Sec.~\ref{sec:Model}, we present
the model and its exact diagonalization. Analytical expressions for
the dynamic free energy and the YLF zeros are determined in Sec.~\ref{sec:Results}
together with numerical results. Our concluding remarks are given
in Sec.~\ref{sec:CONCLUSION}. 

\section{The model\label{sec:Model}}

We consider a free fermion chain with long-range hoppings under twisted
boundary condition given by the Hamiltonian
\begin{equation}
H(\delta)=\sum_{j=1}^{L}\frac{1+\left(-1\right)^{j}\delta}{2}\sum_{\ell=1}^{D}J_{j,2\ell-1}\left(c_{j}^{\dagger}c_{j+2\ell-1}^{\phantom{\dagger}}+{\rm H.c.}\right).\label{eq:H-free-fermion-1}
\end{equation}
We consider systems of $L$ sites in which $L$ is even. The hopping
amplitude decays as { $J_{j,\ell}=J2^{\nu}(\ell+1)^{-\nu}$.}
Here, the constant $J$ sets the energy (or inverse time) unit of
the system (and, from now on, is set to $J=1$), {and
$c_{j+L}=\exp\left(-\phi\pi i\right)c_{j}$ ($1\le j\le L$),} where
$\phi$ defines the type of boundary condition: $\phi=0$ means periodic
boundary condition (PBC) and $\phi=1$ means anti-periodic boundary
condition (APBC). The exponent $\nu\geq0$ controls
the decay of the hopping amplitude with the distance, $D$ is the
hopping range, and $\delta$ is the dimerization parameter which tunes
the system across an equilibrium quantum phase transition (QPT) at
$\delta=0$. For $D=1$, this model recovers the dimerized chain with
nearest-neighbor hopping, also known as Su-Schrieffer-Heeger (SSH)
chain~\citep{SShmodel}. This model, for some particular choice of
the parameters, was used to study symmetry-resolved entanglement entropy~\citep{Ares_Calabrese2022LongRange,LongRangeJones}.
This is an interesting model because it allows one to investigate
the effects of {long-range hopping} and is amenable
to be solved by free-fermion techniques.

{Note that the gauge transformation $c_{j}\rightarrow e^{-i\pi\Phi j/L}c_{j}$
makes the Hamiltonian translational invariant and, thus, can be diagonalized
by the Fourier series.} For the sake of completeness, we present the
main steps below. First, we introduce the new fermionic operators
$\gamma_{q}$ and $\eta_{q}$ by 
\begin{equation}
c_{2j}=\sqrt{\frac{2}{L}}\sum_{q}e^{2iqj}\eta_{q},\mbox{ and }c_{2j-1}=\sqrt{\frac{2}{L}}\sum_{q}e^{iq\left(2j-1\right)}\gamma_{q},\label{eq:Fourier}
\end{equation}
where the momenta are $\mbox{ }q=q_{n}=\frac{2\pi}{aL}\left(n-\frac{\phi}{2}\right)$,
$n=1,2,\dots,L/2$, and, from now on, we set the lattice spacing to
$a=1$. In terms of $\gamma_{q}$ and $\eta_{q}$ the Hamiltonian
is 

\begin{eqnarray}
H & = & \sum_{q}\left(\begin{array}{cc}
\gamma_{q}^{\dagger} & \eta_{q}^{\dagger}\end{array}\right)\left(\begin{array}{cc}
0 & C_{q}-i\delta S_{q}\\
C_{q}+i\delta S_{q} & 0
\end{array}\right)\left(\begin{array}{c}
\gamma_{q}^{\phantom{\dagger}}\\
\eta_{q}^{\phantom{\dagger}}
\end{array}\right),\nonumber \\
 & = & \sum_{q}\omega_{q,\delta}\left(\alpha_{+,q,\delta}^{\dagger}\alpha_{+,q,\delta}^{\phantom{\dagger}}-\alpha_{-,q,\delta}^{\dagger}\alpha_{-,q,\delta}^{\phantom{\dagger}}\right),\label{eq:H-diag}
\end{eqnarray}
where 
\begin{eqnarray}
C_{q}=C_{q}(\nu,D) & = & \sum_{\ell=1}^{D}\ell^{-\nu}\cos\left((2\ell-1)q\right),\label{eq:C}\\
S_{q}=S_{q}(\nu,D) & = & \sum_{\ell=1}^{D}\ell^{-\nu}\sin\left((2\ell-1)q\right),\label{eq:S}\\
\omega_{q,\delta}=\omega_{q,\delta}(\nu,D) & = & \sqrt{C_{q}^{2}+\delta^{2}S_{q}^{2}},\label{eq:dispersion}
\end{eqnarray}
 and 
\begin{equation}
\alpha_{\pm,q,\delta}=\frac{1}{\sqrt{2}}\left(e^{i\theta_{q,\delta}}\gamma_{q}\pm e^{-i\theta_{q,\delta}}\eta_{q}\right)\label{eq:alfa-beta-gama-eta}
\end{equation}
are the eigen-operators associated to positive and negative branches
of the dispersion relation $\pm\omega_{q,\delta}(\nu,D)$. Here, $\cos2\theta_{q,\delta}=C_{q}/\omega_{q,\delta}$
and $\sin2\theta_{q,\delta}=\delta S_{q}/\omega_{q,\delta}$.\footnote{The quantities defined in Eqs.~(\ref{eq:C})--(\ref{eq:alfa-beta-gama-eta})
depend on $q,$ $\delta$, $\nu$ and $D$. To lighten the notation,
only the dependence of $q$ and $\delta$ is kept in the subscript.} 

Finally, notice that

\begin{equation}
C_{\frac{\pi}{2}-q}=-C_{\frac{\pi}{2}+q},\mbox{ and }S_{\frac{\pi}{2}-q}=S_{\frac{\pi}{2}+q},\label{eq:CS-2}
\end{equation}
 and that the ground state of $H(\delta)$ is 
\begin{equation}
\left|\psi_{0}(\delta)\right\rangle =\prod_{q}\alpha_{-,q,\delta}^{\dagger}\left|0\right\rangle ,
\end{equation}
 where the product is over all $q$'s in Eq.~(\ref{eq:Fourier}).

It is worth mentioning that for some special values of $\nu$ and
$D$, the functions $C_{q}$ and $S_{q}$ can also be written in terms
of some well known functions, as depicted in Table~\ref{tab1}. For
$\nu=\infty$ or $D=1$, Eq.~(\ref{eq:dispersion}) recovers that
of the nearest-neighbor hopping problem $\omega_{q,\delta}=\sqrt{\cos^{2}(q)+\delta^{2}\sin^{2}(q)}$.
The case $\nu=0$ and $D=L/4$ is very peculiar and presents some
anomalous characteristics (see Appendix~\ref{AP-A}): (i) The ground-state
energy $E_{0}\sim aL\ln L+bL$ is not extensive. (ii) Different boundary
conditions lead to distinct behaviors. For PBC (APBC), the system
is gapless (gapped) at half filling. In addition, the difference $E_{0}^{\text{APBC}}-E_{0}^{\text{PBC}}\sim-L\ln L$.

\begin{table*}
\begin{centering}
\caption{The functions $C_{q}(\nu,D)$ and $S_{q}(\nu,D)$ for some special
values of $\nu$ and $D$. $\text{Li}_{\nu}(z)$ is the polylogarithm
function of order $\nu$.\label{tab1}}
\par\end{centering}
\centering{}%
\begin{tabular}{ccccc}
\hline 
 &  & $C_{q}(\nu,D)$ &  & $S_{q}(\nu,D)$\tabularnewline
\hline 
\hline 
$\nu=0$ &  & $\frac{\sin\left(2Dq\right)}{2\sin q}$ &  & $\frac{1-\cos\left(2Dq\right)}{2\sin(q)}$\tabularnewline
$\nu=1$ $D=\infty$ &  & $-1/2\left[\cos\left(q\right)\ln\left(\left|2\sin q/2\right|\right)-\sin\left(q\right)(\pi-2q)\right]$ &  & $1/2\left[\cos\left(q\right)(\pi-2q)+\sin\left(q\right)\ln\left(\left|2\sin q/2\right|\right)\right]$\tabularnewline
$\nu=2$ $D=\infty$ &  & $\frac{\pi^{2}}{6}-\frac{2q}{2}+\frac{q^{2}}{4}$ &  & $-\int_{0}^{q}\ln\left(2\sin(t/2)\right)dt$\tabularnewline
$\nu=\infty$ &  & $\cos\left(q\right)$ &  & $\sin\left(q\right)$\tabularnewline
$D=\infty$ &  & $\text{Re}\left(e^{-iq}\text{Li}_{\nu}\left(e^{i2q}\right)\right)$ &  & $\text{Im}\left(e^{-iq}\text{Li}_{\nu}\left(e^{i2q}\right)\right)$\tabularnewline
\hline 
\end{tabular}
\end{table*}

\section{Results\label{sec:Results} }

\subsection{The dynamic free energy and the YLF zeros}

As we already mentioned, in our quench protocol the system is initialized
in $\left|\psi_{0}(\delta_{0})\right\rangle $, the ground state of
$H(\delta_{0})$, and time-evolved according to $H(\delta)$. Only
$\delta$ is changed in the sudden quench, $\nu$ and $D$ remain
constants. The return probability amplitude $Z(t)=\left\langle \psi_{0}(\delta_{0})\left|e^{-iH(\delta)t}\right|\psi_{0}(\delta_{0})\right\rangle $
can be evaluated following the same procedure of Ref.~\citep{NetoRafaelXavierPRBL2022,NetoRafaelXavierunpublish}.
For completeness, we present below the main steps. 

To time-evolve $\left|\psi_{0}(\delta_{0})\right\rangle $, we need
the relation between the pre- and post-quench eigen-operators $\alpha_{\pm,q,\delta_{0}}$
and $\alpha_{\pm,q,\delta}$ {[}see Eq.~(\ref{eq:H-diag}){]}. This
task is simple, since the wavenumbers $q$ in (\ref{eq:Fourier}),
and, therefore, $\gamma_{q}$ and $\eta_{q}$, do not depend on $\delta$.
Then, from Eq.~(\ref{eq:alfa-beta-gama-eta}), we find that 
\begin{equation}
\alpha_{-,q,\delta_{0}}^{\dagger}=\cos\left(\Delta\theta_{q,\delta,\delta_{0}}\right)\alpha_{-,q,\delta}^{\dagger}+i\sin\left(\Delta\theta_{q,\delta,\delta_{0}}\right)\alpha_{+,q,\delta}^{\dagger},
\end{equation}
where $\Delta\theta_{q,\delta,\delta_{0}}=\theta_{q,\delta}-\theta_{q,\delta_{0}}$.
Therefore, 
\begin{eqnarray}
Z(t) & = & \left\langle 0\left|\prod_{q}\alpha_{-,q,\delta_{0}}e^{-iHt}\prod_{k}\alpha_{-,k,\delta_{0}}^{\dagger}\right|0\right\rangle \nonumber \\
 & = & \prod_{q}\left[\cos\left(\omega_{q,\delta}t\right)+ig_{q,\delta,\delta_{0}}\sin\left(\omega_{q,\delta}t\right)\right],\label{eq:Zt}
\end{eqnarray}
where $0\leq g_{q,\delta,\delta_{0}}=\frac{C_{q}^{2}+\delta\delta_{0}S_{q}^{2}}{\omega_{q,\delta}\omega_{q,\delta_{0}}}\leq1$. 

Finally, the dynamic free energy $f(t)\equiv-L^{-1}\ln\left|Z(t)\right|^{2}$
is 

\begin{equation}
f(t)=-\frac{1}{L}\sum_{q}\ln\left[\cos^{2}\left(\omega_{q,\delta}t\right)+g_{q,\delta,\delta_{0}}^{2}\sin^{2}\left(\omega_{q,\delta}t\right)\right],\label{eq:DynEner}
\end{equation}
and, in the thermodynamic limit, we can replace the sum by the integral

\begin{equation}
f(t)=-\intop_{0}^{\pi/2}\frac{dq}{\pi}\ln\left[\cos^{2}\left(\omega_{q,\delta}t\right)+g_{q,\delta,\delta_{0}}^{2}\sin^{2}\left(\omega_{q,\delta}t\right)\right],\label{eq:DynEner-1}
\end{equation}
 where the properties (\ref{eq:CS-2}) where used to shorten the integration
limit. 

Let $\zeta_{n,m}=t_{n,m}+i\tau_{n}$ (or $\zeta_{q_{n},m}=t_{q_{n},m}+i\tau_{q_{n}}$)
be the YLF zeros of $Z$. From Eq.~(\ref{eq:Zt}), it is simple to
show that 
\begin{equation}
\zeta_{q_{n},m}=\frac{\left(m-\frac{1}{2}\right)\pi+\frac{i}{2}\ln\left(\frac{1+g_{q_{n},\delta,\delta_{0}}}{1-g_{q_{n},\delta,\delta_{0}}}\right)}{\omega_{q_{n},\delta}},\label{eq:YLF-zeros}
\end{equation}
 were $m\in\mathbb{N}_{+}$ is the $m$th accumulation line of YLF
zeros, and $n=1,\dots,\frac{L}{2}$ labels the $n$th wavenumber $q_{n}$
in (\ref{eq:Fourier}). Although there are $\frac{L}{2}$ YLF zeros
per accumulation line, not all of them are distinct because of (\ref{eq:CS-2}).
For $\frac{L}{2}$ odd, there are $\frac{1}{2}\left(\frac{L}{2}-1\right)$
distinct zeros (which are doubly degenerated), and one (for $q=\pi$
for PBC and $q=\frac{\pi}{2}$ for APBC) has $\left|\tau_{q_{n}}\right|=\infty$.
Thus, effectively there are $\frac{1}{2}\left(\frac{L}{2}-1\right)$
zeros. For PBC and $\frac{L}{2}$ even, there are $\frac{1}{2}\left(\frac{L}{2}-2\right)$
distinct zeros (which are doubly degenerated), and 2 zeros (for $q=\frac{\pi}{2}$
and $\pi$) with $\left|\tau_{q_{n}}\right|=\infty$. For APBC and
$\frac{L}{2}$ even, there are $\frac{L}{4}$ doubly degenerated distinct
zeros. 

The DQPTs occur whenever $\tau_{q_{n}}=0$ and, thus, from Eq.~(\ref{eq:YLF-zeros}),
they can only happen if $C_{q^{c}}^{2}+\delta\delta_{0}S_{q^{c}}^{2}=0$,
i.e., 
\begin{equation}
T_{q^{c}}^{2}(\nu,D)\equiv\frac{S_{q^{c}}^{2}(\nu,D)}{C_{q^{c}}^{2}(\nu,D)}=-\frac{1}{\delta\delta_{0}}.\label{eq:T}
\end{equation}
Notice the necessary condition $\delta\delta_{0}<0$ which corresponds
to the quench crossing the equilibrium QPT of the model at $\delta_{\text{eq}}^{c}=0$.\footnote{$\delta_{\text{eq}}^{c}$ is determined by requiring $\omega_{q^{c}}=0$
in Eq.~(\ref{eq:dispersion}). For $\nu\ne0$, $S_{q}(\nu,D)\ne0\ \forall q$,
and thus, $\delta_{\text{eq}}^{c}=0$. Notice it does not depend on
$\nu$ which is quite different from the transverse-field Ising model
with long-range interaction~\citep{DynamicLongRangePRB2017}. } Once the set $\left\{ q^{c}\right\} $ is determined from Eq.~(\ref{eq:T}),
the time instants of the DQPTs are simply $t_{\left\{ q^{c}\right\} ,m}^{c}=\frac{\left(2m-1\right)\pi}{2\omega_{q^{c},\delta}}$.
Due to the properties (\ref{eq:CS-2}), if $q^{c}$ is a solution
of (\ref{eq:T}), so is $\pi-q^{c}$. In addition, they provide the
same YLF zero since $\omega_{q,\delta}=\omega_{\pi-q,\delta}$. Thus,
it is sufficient to consider only the values of $q$ in the domain
$\left[0,\frac{\pi}{2}\right]$ when solving for $\left\{ q^{c}\right\} $
in (\ref{eq:T}).

In general, Eq.~(\ref{eq:T}) admits no solution for finite systems
since $\left\{ q_{n}\right\} $ in Eq.~(\ref{eq:Fourier}) is a discrete
set. Nonetheless, as reported in Appendix~\ref{AP-A}, for some special
values of $\nu$, $D$, $\delta$ and $\delta_{0}$, Eq.~(\ref{eq:T})
admits solutions for finite systems and, thus, for a real-time instant
$t^{c}$, $f(t^{c})$ is non-analytic even for finite $L$. Non-analyticities
in finite-size systems were also reported in Refs.~\onlinecite{DynamSirkerPRB2014,KarraschPhysRevB.87.195104}. 

\subsection{The case of nearest-neighbor ($D=1$) and third-nearest-neighbor
($D=2$) hoppings }

For completeness, we now briefly review the results for $D=1$ and
compare them with the case $D=2$. It turns out that this comparison
is very instructive to understand the case of generic $D$.

\begin{figure}
\begin{centering}
\includegraphics[clip,width=0.85\columnwidth]{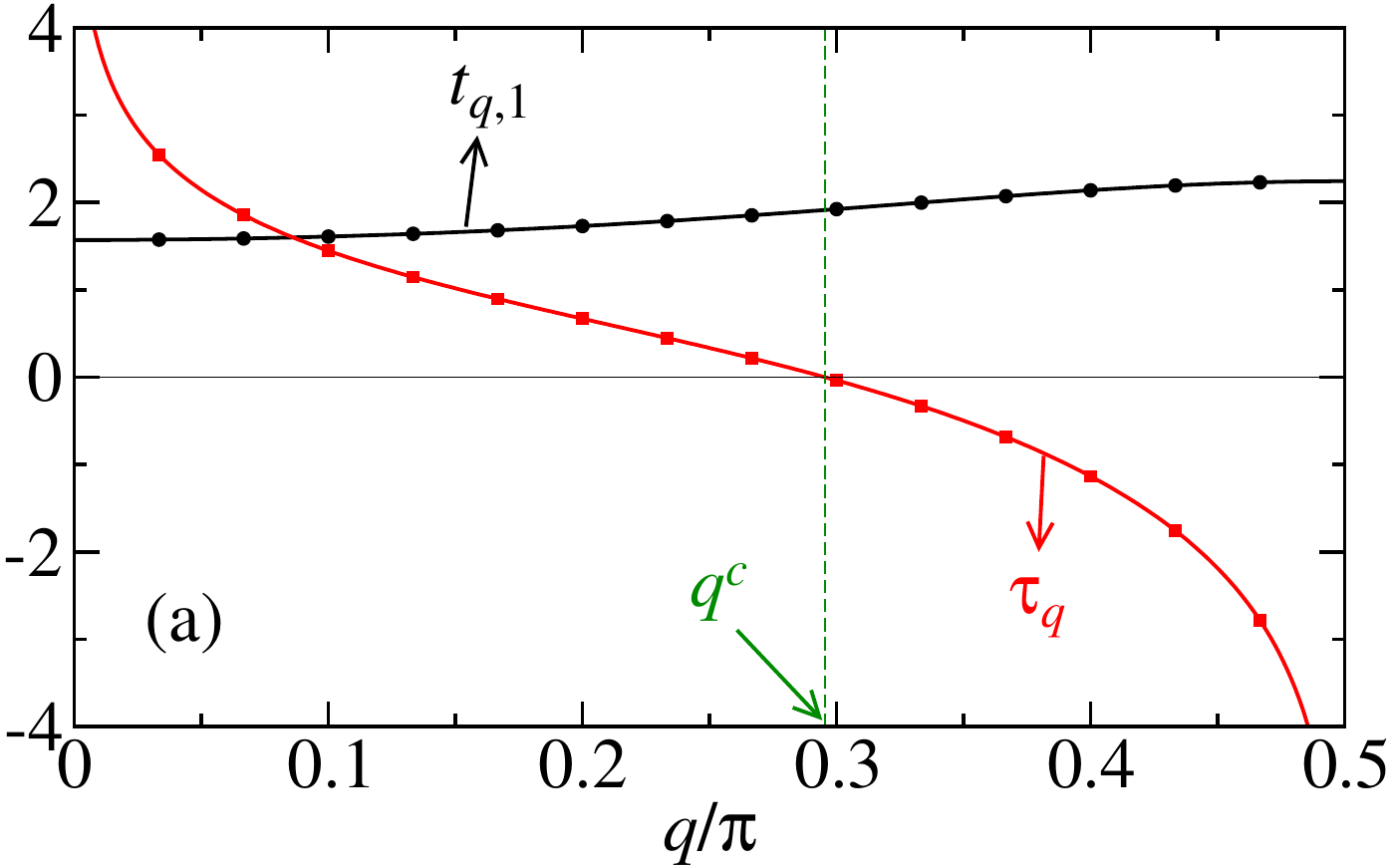}\\
\includegraphics[clip,width=0.85\columnwidth]{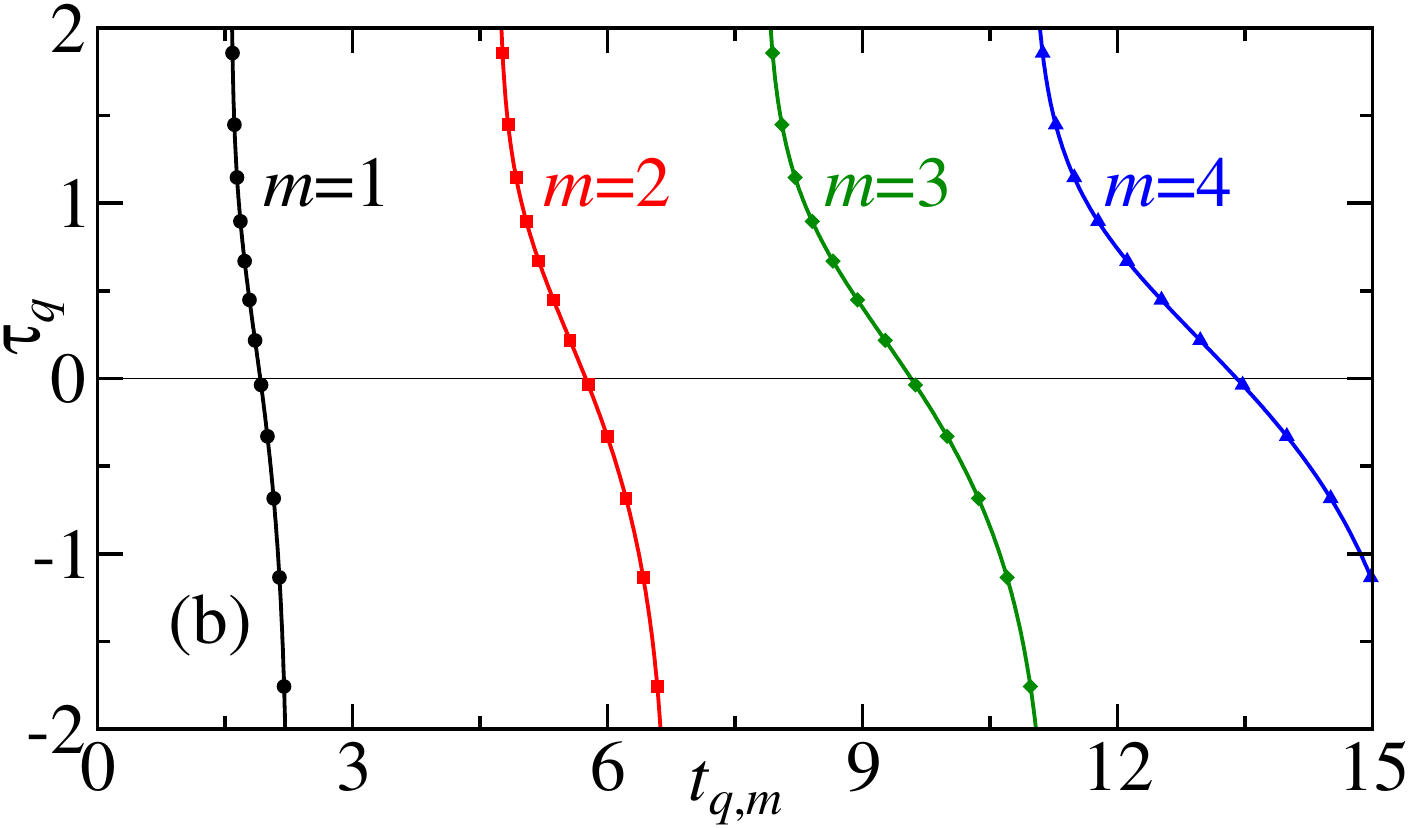}\\
\includegraphics[clip,width=0.85\columnwidth]{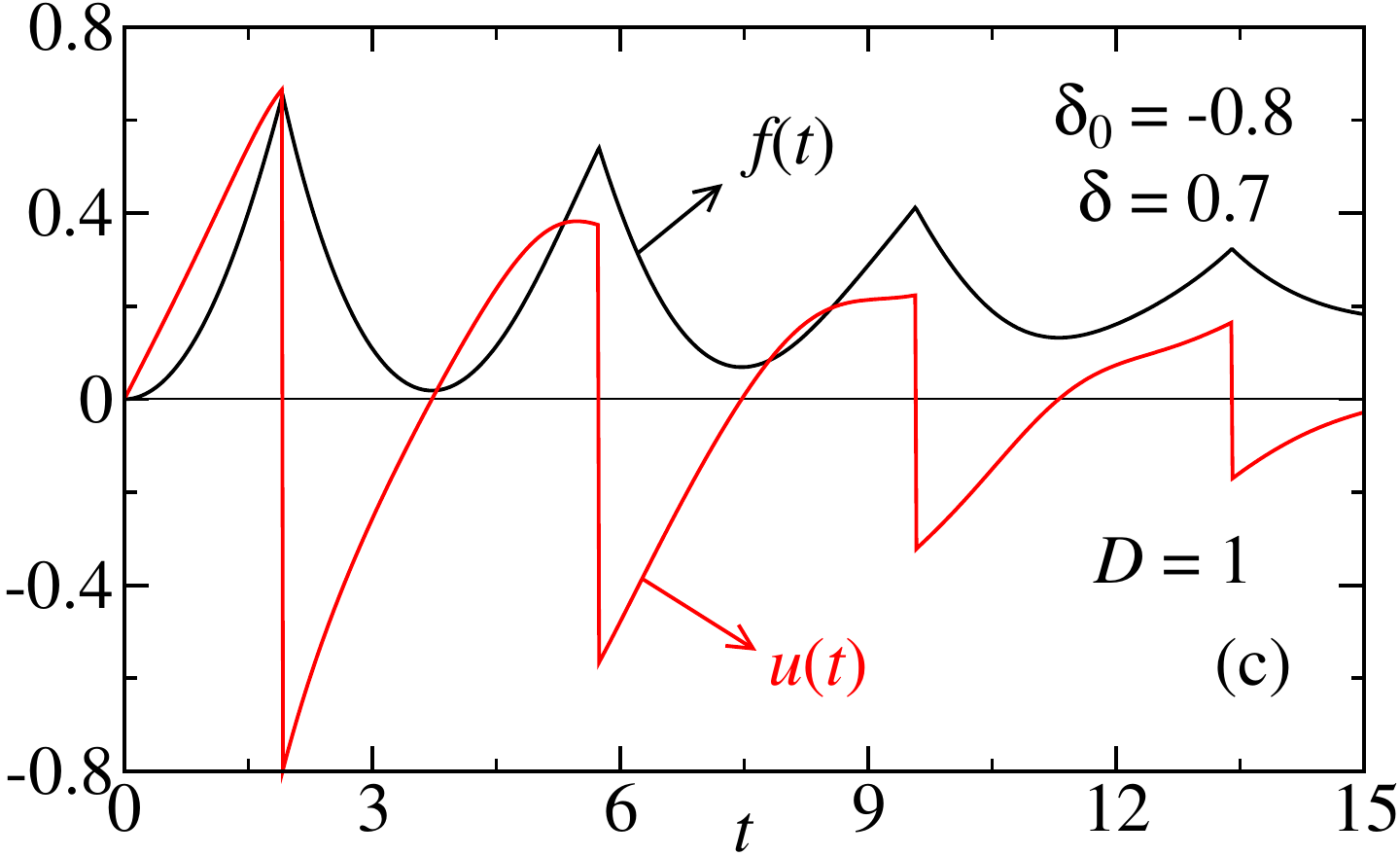}
\par\end{centering}
\caption{In (a) and (b) we show the Yang-Lee-Fisher zeros $\zeta_{q_{n},m}$
{[}Eq.~(\ref{eq:YLF-zeros}){]} for $D=1$. (a) $t_{q,1}$ and $\tau_{q}$
as a function of the momentum $q$ (for accumulation line $m=1$).
(b) The complex-time plane (for $m=1,\dots,4$). (c) The associated
dynamical free energy $f$ {[}Eq.~(\ref{eq:DynEner-1}){]} and its
time derivative $u$ as a function of $t$. In (a), $q^{c}$ denotes
the momentum associated with the real-time YLF zero. The symbols in
(a) and (b) correspond to the allowed values of $q$ for a finite
chain of length $L=60$. The continuous lines in (a), (b) and (c)
correspond to the thermodynamic limit. The sudden quench is from $\delta_{0}=-0.8$
to $\delta=0.7$. \label{fig:D1} }

\end{figure}

For $D=1$ and $\delta\delta_{0}<0$, Eq.~(\ref{eq:T}) gives a single
solution $q^{c}=\arctan\left(\frac{1}{\sqrt{-\delta\delta_{0}}}\right)\in\left[0,\frac{\pi}{2}\right]$
{[}see Fig.~\hyperref[fig:D1]{\ref{fig:D1}(a)}{]}. This means that
each accumulation line in (\ref{eq:YLF-zeros}) provides only one
real-time instant $t_{q^{c},m}=\left(m-\frac{1}{2}\right)\pi\sqrt{\frac{1-\delta\delta_{0}}{\delta\left(\delta-\delta_{0}\right)}}$
in which the dynamic free energy is non-analytic in the thermodynamic
limit {[}see Fig.~\hyperref[fig:D1]{\ref{fig:D1}(b)}{]}. This non-analyticity
is manifest as a cusp in $f\left(t\right)$ (or a discontinuity in
the dynamic internal energy $u\equiv\frac{\partial f}{\partial t}$)
at $t=\zeta_{q^{c},m}=t_{q^{c},m}$ {[}see Fig.~\hyperref[fig:D1]{\ref{fig:D1}(c)}{]}.
Note that for the case $D=1$, the results do not depend on the value
of $\nu$.

Precisely, the non-analyticity of the dynamic free energy can be quantified
by analyzing the behavior of the YLF zeros near the real-time axis.
From the Weierstrass factorization theorem~\citep{YLzerosI,YLzerosII,bookFisherYLzeros},
the singular part of the free energy due to the zero in the $m$th
accumulation line is $f_{\text{n-a}}=-L^{-1}\sum_{n}\ln\left(\zeta-\zeta_{q_{n},m}\right)+\text{c.c.}$
. Here, c.c. stands for complex conjugate and accounts for the zeros
of $\bar{Z}$ (the complex conjugate of $Z$). In the thermodynamic
limit, $\zeta_{q_{n},m}$ in Eq.~(\ref{eq:YLF-zeros}) can be expanded
near $q^{c}$ {[}see Fig.~\hyperref[fig:D1]{\ref{fig:D1}(a)}{]}.
Then, the real-time non-analyticity of the dynamic internal energy
is quantified by 
\begin{equation}
u_{\text{n-a}}=-\frac{1}{\pi}\int_{-\delta q}^{\delta q}\frac{d\tilde{q}}{\Delta t-\left(A_{q^{c},m}-iB_{q^{c}}\right)\tilde{q}}+\text{c.c.},\label{eq:u-D1}
\end{equation}
where $\Delta t=t-\zeta_{q^{c},m}=t-t_{q^{c},m}$, $\tilde{q}=q-q^{c}$,
$A_{q^{c},m}=\left.\frac{\partial t_{q,m}}{\partial q}\right|_{q=q^{c}}=-\frac{\delta_{0}\left(1-\delta^{2}\right)}{\left(\delta-\delta_{0}\right)\sqrt{-\delta\delta_{0}}}t_{q^{c},m}$,
$B_{q^{c}}=-\left.\frac{\partial\tau_{q}}{\partial q}\right|_{q=q^{c}}=2\left|\delta\right|\omega_{q^{c},\delta}^{-3}$,
and $\delta q$\textcolor{red}{{} }is a positive constant whose value
is unimportant for quantifying the non-analyticity of $u$. The numerical
prefactor is $\pi^{-1}$ and not $\left(2\pi\right)^{-1}$ because
we are using the properties (\ref{eq:CS-2}) to take into account
the other YLF zero in the interval $q\in\left[\frac{\pi}{2},\pi\right]$.
By a simple integration (via residues, for instance), we can show
that first derivative of the dynamic free energy has a discontinuity
given by
\begin{equation}
\Delta u\left(\zeta=\zeta_{q,m}\right)=\lim_{\Delta t\rightarrow0^{+}}u_{\text{n-a}}-\lim_{\Delta t\rightarrow0^{-}}u_{\text{n-a}}=-\frac{4B_{q^{c}}}{A_{q^{c},m}^{2}+B_{q^{c}}^{2}}.
\end{equation}
This is because the pole at $\tilde{q}=\frac{\Delta t}{A_{q^{c},m}-iB_{q^{c}}}$
crosses the real-$\tilde{q}$ axis when $\Delta t$ changes sign.
We have confirmed this result via numerical integration of (\ref{eq:DynEner-1}).

\begin{figure}
\begin{centering}
\includegraphics[clip,width=0.85\columnwidth]{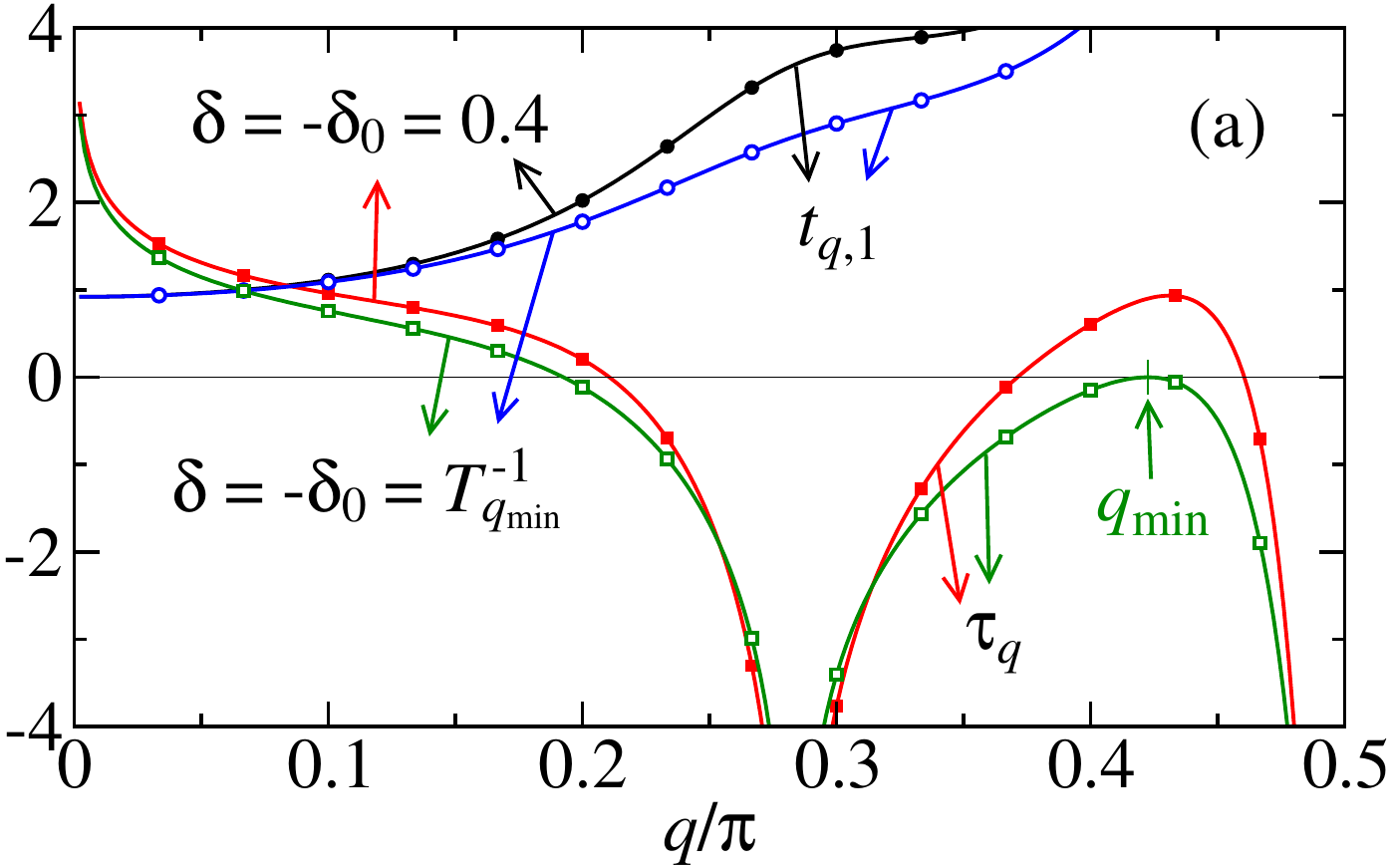}\\
\includegraphics[clip,width=0.85\columnwidth]{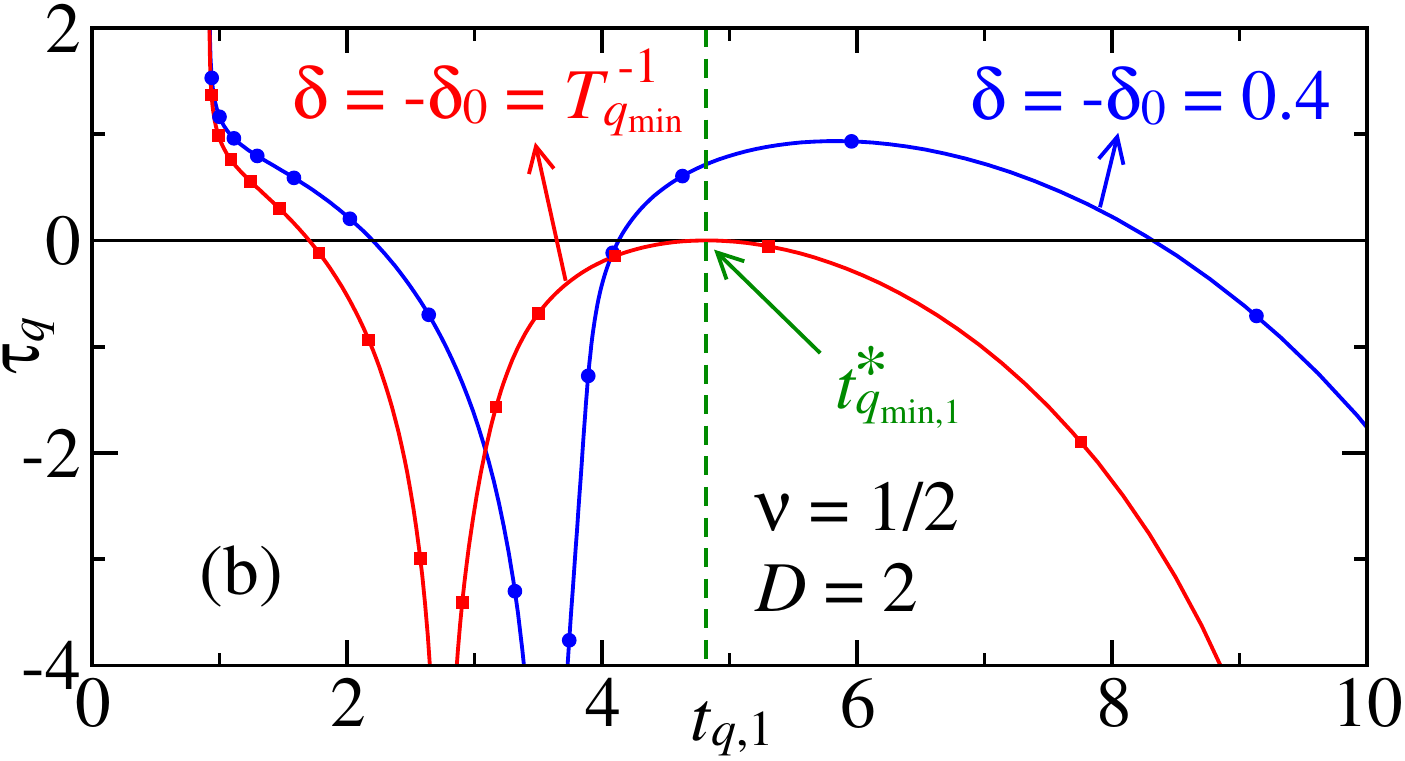}\\
\includegraphics[clip,width=0.85\columnwidth]{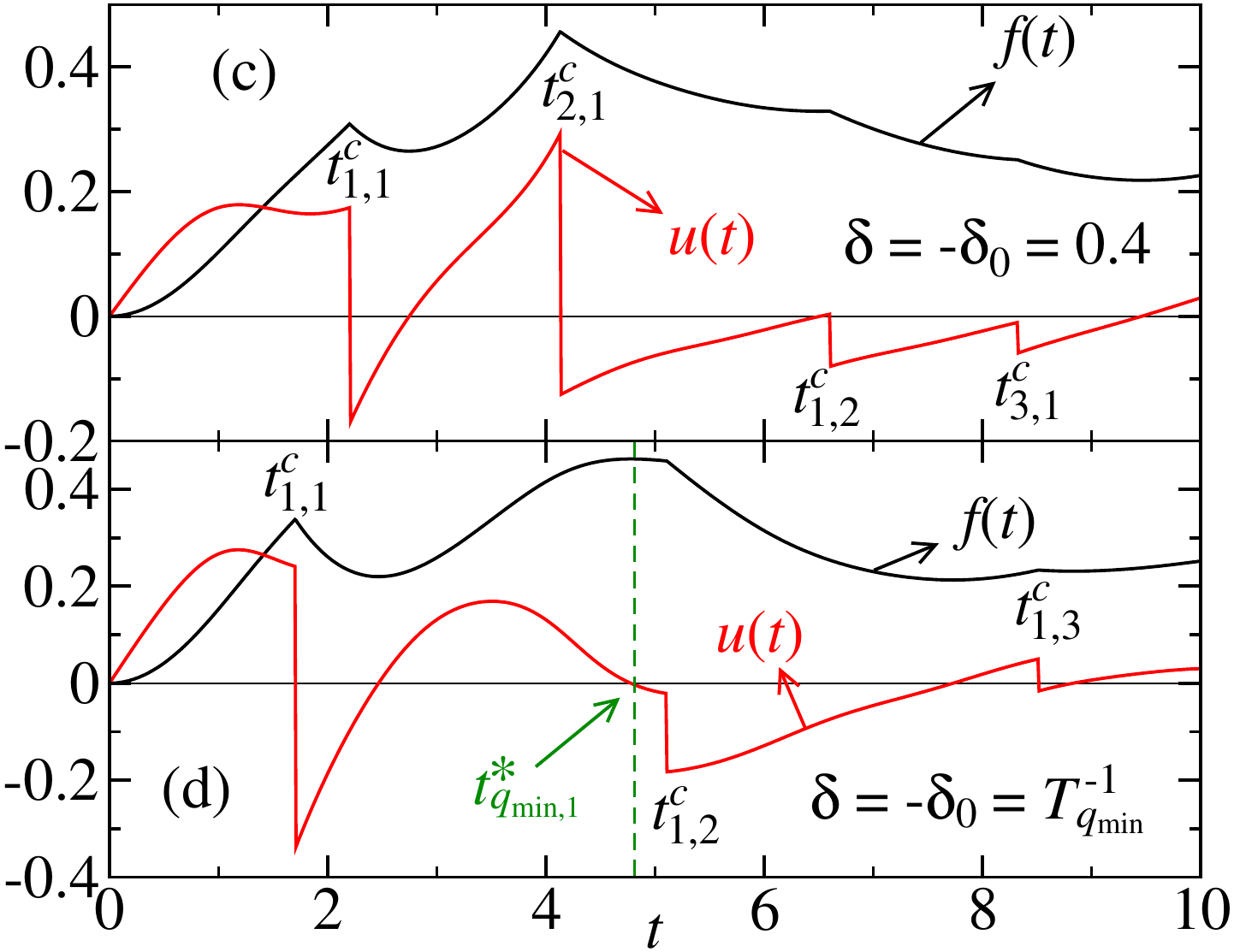}
\par\end{centering}
\caption{\label{fig:D2} In (a) and (b) we show the Yang-Lee-Fisher zeros $\zeta_{q_{n},m}$
{[}Eq.~(\ref{eq:YLF-zeros}){]} for $D=2$ and $\nu=1/2$. (a) $t_{q,1}$
and $\tau_{q}$ as a function of the momentum $q$ (for accumulation
line $m=1$) and (b) in the complex-time plane (for $m=1$). (c) and
(d) The associated dynamical free energy $f$ {[}Eq.~(\ref{eq:DynEner-1}){]}
and its time derivative $u$ as a function of $t$. We consider two
sudden quenches: (c) from $\delta_{0}=-0.4$ to $\delta=0.4$ (fulfilling
$-\frac{1}{\delta\delta_{0}}>T_{q_{\text{min}}}^{2}$) and (d) from
$\delta_{0}=-T_{q_{\text{min}}}^{-1}\left(\nu,D\right)$ to $\delta=T_{q_{\text{min}}}^{-1}\left(\nu,D\right)\approx0.52$
(see legends). In (a), $q_{\text{min}}$ denotes the momentum associated
with the real-time YLF zero which only touches the real-time axis.
In (b) and (d), the corresponding time instant $t_{q_{\text{min}}}$
is highlighted. The symbols in (a) and (b) correspond to the allowed
values of $q$ for a finite chain of length $L=60$. The continuous
lines in (a), (b), (c) and (d) correspond to the thermodynamic limit.}
\end{figure}

For $D=2$ and $\delta\delta_{0}<0$, the situation is more involved.
If $v<\log_{2}3\approx1.585$, Eq.~(\ref{eq:T}) admits two additional
solutions if $-\frac{1}{\delta\delta_{0}}>T_{q_{\text{min}}}^{2}(\nu,D)=\left(\frac{3+2^{\nu}}{3-2^{\nu}}\right)^{3}\left(\frac{2^{\nu}+1}{2^{\nu}-1}\right)$
{[}see Fig.~\hyperref[fig:D2]{\ref{fig:D2}(a)}{]}. This is because
$T_{q}^{2}(\nu,D)$ has a local minimum at $q_{\text{min}}=\frac{1}{2}\arccos\left(-\frac{4^{\nu}+3}{2^{2+\nu}}\right)$.
Thus, each accumulation line of YLF zeros crosses the real-time axis
at three different instants {[}see Fig.~\hyperref[fig:D2]{\ref{fig:D2}(b)}{]}.
The corresponding density of YLF zeros crossing the real-time axis
is a constant. Hence, as in the case $D=1$, the corresponding non-analyticities
are cusps in $f(t)$ at those time instants {[}see Figs.~\hyperref[fig:D2]{\ref{fig:D2}(c)}
and \hyperref[fig:D2]{(d)}{]}. 

However, it is not straightforward to anticipate the resulting singularity
when the two additional YLF zeros become degenerate, i.e., when $-\frac{1}{\delta\delta_{0}}=T_{q_{\text{min}}}^{2}(\nu,D)$.
Following the same steps as in Eq.~(\ref{eq:u-D1}), the singular
part of the dynamical internal energy around the time instant $\zeta_{q_{\text{min}},m}=t_{q_{\text{min}},m}^{*}$
is 
\begin{equation}
u_{\text{n-a}}=-\frac{1}{\pi}\int_{-\delta q}^{\delta q}\frac{d\tilde{q}}{\Delta t-\left(A_{q_{\text{min}},m}\tilde{q}-\frac{i}{2}C_{q_{\text{min}}}\tilde{q}^{2}\right)}+\text{c.c.}
\end{equation}
where $\Delta t=t-\zeta_{q_{\text{min}},m}=t-t_{q_{\text{min}},m}^{*}$,
$\tilde{q}=q-q_{\text{min}}$, $A_{q_{\text{min}},m}=\left.\frac{\partial t_{q,m}}{\partial q}\right|_{q=q_{\text{min}}}$,
$C_{q_{\text{min}}}=-\left.\frac{\partial^{2}\tau_{q}}{\partial q^{2}}\right|_{q=q_{\text{min}}}$,
and $\delta q$, as before, is an unimportant positive constant. As
for the case $D=1$, the non-analytical behavior of $u_{\text{n-a}}$
comes when a pole crosses the real-$\tilde{q}$ axis. However, we
now face the situation where the integrand of $u_{\text{n-a}}$ has
two poles. It is easy to see that one of the poles always remains
far from the real-$\tilde{q}$ axis and, thus, does not contribute
to the non-analyticity. The other one does not cross the real-$\tilde{q}$
axis either. It only touches it when $\Delta t=0$. As a result, the
limit $u_{\text{n-a}}(t)$ as $t\rightarrow t_{q_{\text{min}}}$ exists,
i.e., $\Delta u=\lim_{\Delta t\rightarrow0^{+}}u_{\text{n-a}}-\lim_{\Delta t\rightarrow0^{-}}u_{\text{n-a}}=0$.
The same reasoning applies to all derivatives of $u$. Finally, we
conclude that although $f$ is non-analytic at $t_{q_{\text{min}}}$,
it is a smooth function (all derivatives exist) at that time instant
{[}see Fig.~\hyperref[fig:D2]{\ref{fig:D2}(d)}{]}. Nonetheless,
we recall that this non-analyticity poses a numerical challenge in
computing $f$ and its derivatives at that time instant. 

In analogy to the Ehrenfest's classification of the order of the equilibrium
phase transitions~\citep{Ehrenfest}, we could classify the order
of the DQPTs by the lowest derivative of the dynamic free energy that
is discontinuous at the transition. With this classification in mind,
we observe that when the YLF zeros are not degenerate, the DQPT is
of first order. On the other hand, when the YLF zeros become degenerate,
the DQPT is of infinite order. It is then tempting to state that this
is the dynamic analog of the Berezinskii-Kosterlitz--Thouless (BKT)
transition of equilibrium systems. However, BKT transition has a continuous
of YLF zeros in one of the phases. Here, there is no continuous distribution
of YLF zeros after or before the instant of non-analyticity $t_{q_{\text{min}},m}^{*}$. 

\subsection{Numerical results}

As we show below, this feature of two dynamical QPTs becoming degenerate
(either by fine-tuning $\delta\delta_{0}$ or $\nu$) and the associated
cusps annihilating each other is a general feature for all other values
of the hopping range $D$. 

\begin{figure}[t]
\begin{centering}
\includegraphics[clip,width=0.85\columnwidth]{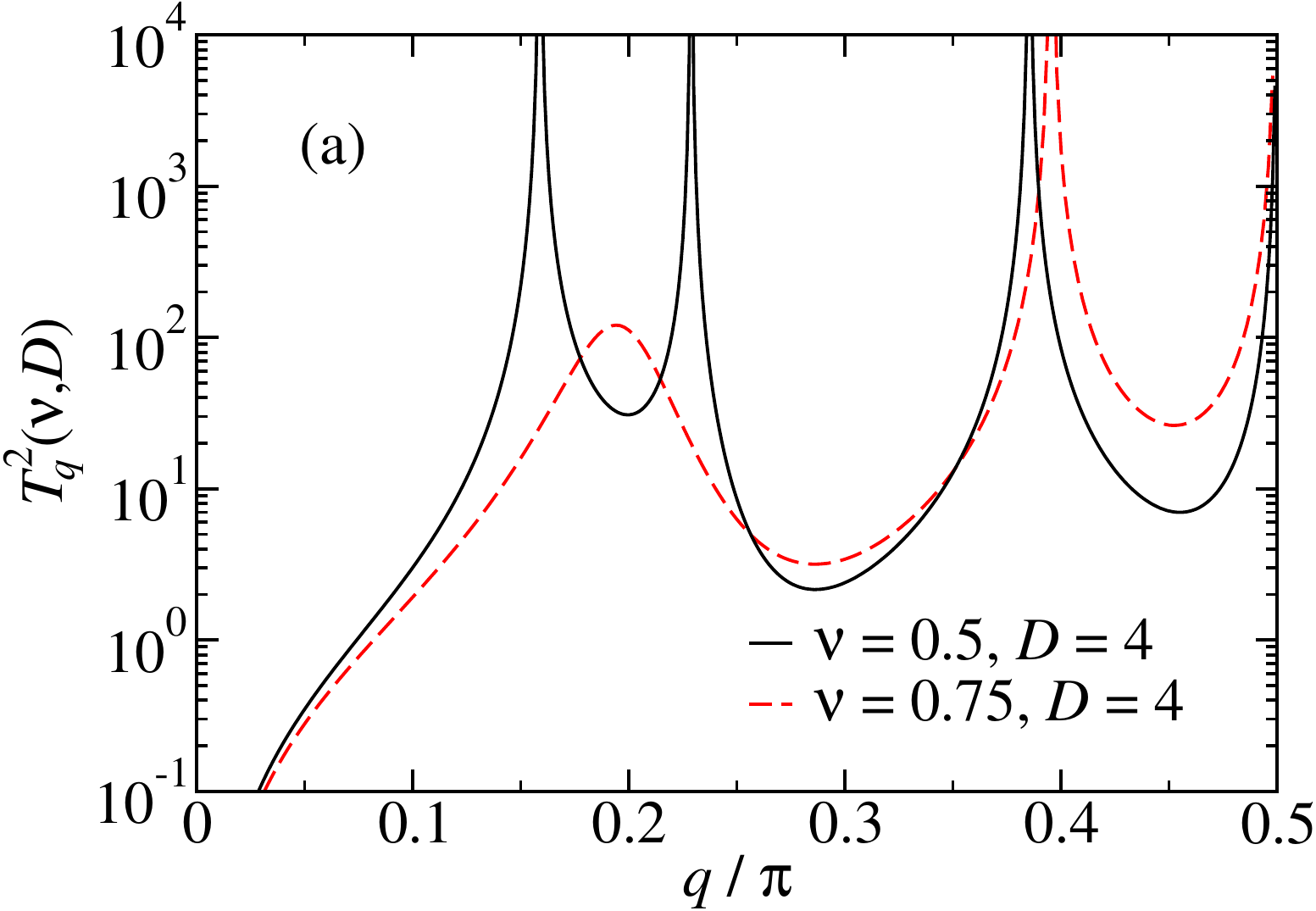}\\
\includegraphics[clip,width=0.8\columnwidth]{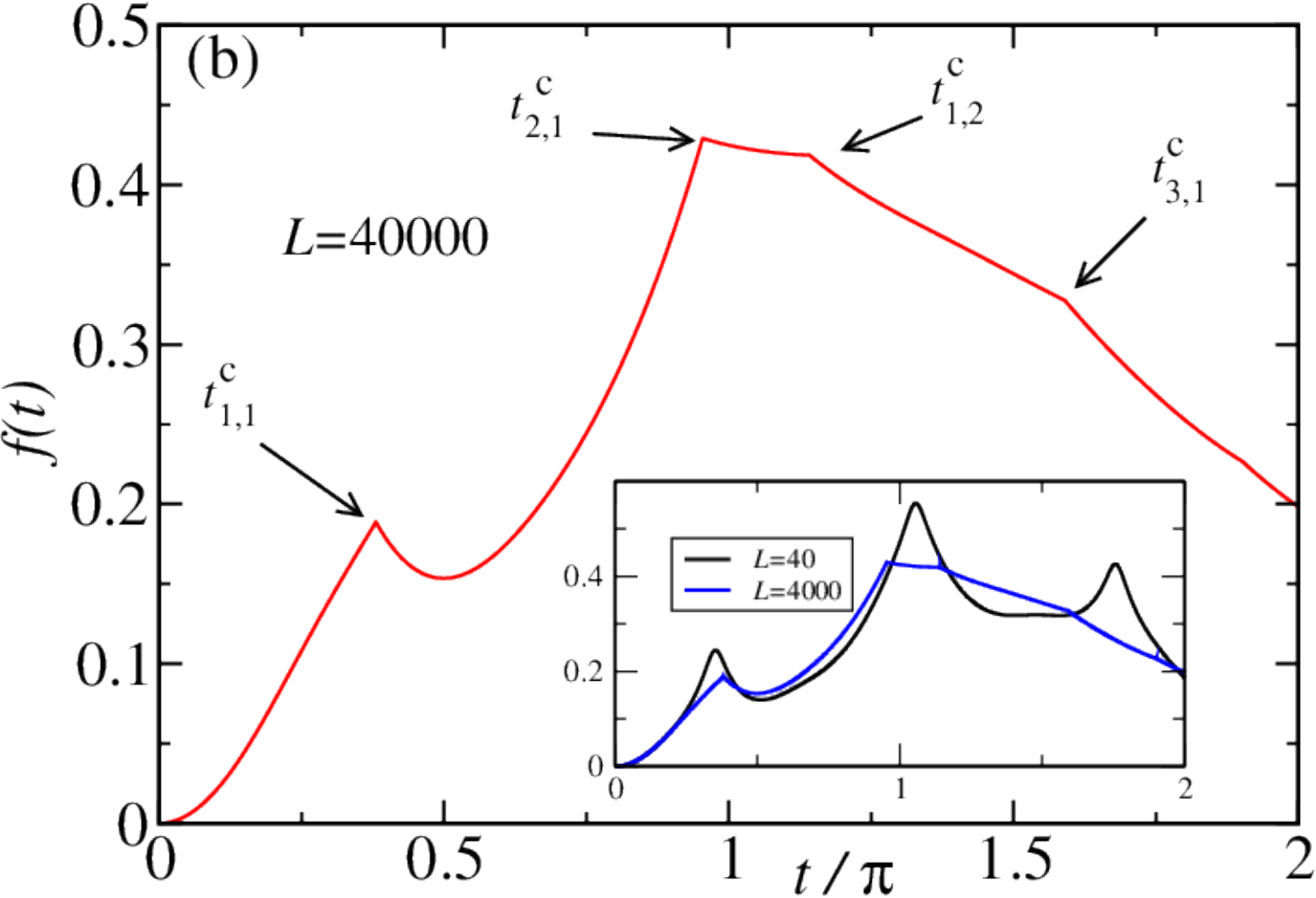}
\par\end{centering}
\caption{(a) $T_{\nu,D}^{2}$ vs. $q$ for $D=4$ and $\nu=0.5$ and $0.75$.
(b) The dynamic free energy $f(t)$ vs. $t$ for a system of size
$L=40\,000$, $D=4$, $\nu=0.5$, and $\delta_{0}=-\text{\ensuremath{\delta=0.5}}$
{[}meaning Eq.~(\ref{eq:T}) has three solutions for $0<q<\frac{\pi}{2}${]}.
The arrows indicate the cusp positions, which are located at $t_{k,m}^{c}$
(see text). Inset: $f(t)$ for $L=40$ and $L=4000$.\label{fig:T-f} }
\end{figure}

We plot in Fig.~\hyperref[fig:T-f]{\ref{fig:T-f}(a)} $T_{q}^{2}(\nu,D)\equiv\frac{S_{q}^{2}(\nu,D)}{C_{q}^{2}(\nu,D)}$
{[}see Eqs.~(\ref{eq:C}) and (\ref{eq:S}){]} for $\nu=0.5$ and
$D=4$. Notice that $T_{\nu,D}^{2}$ diverges for $q$'s such that
$C_{q}(\nu,D)=0$. When $\nu$ is sufficiently small, $T_{\nu,D}^{2}$
has $D-1$ local minima in the domain $q\in\left[0,\frac{\pi}{2}\right]$.
This means that, for sufficiently large $-\frac{1}{\delta\delta_{0}}$,
there are $2D-1$ solutions of Eq.~(\ref{eq:T}). Let $\left\{ q_{k}^{c}\right\} $
be the set of solutions of Eq.~(\ref{eq:T}) for generic values $-\frac{1}{\delta\delta_{0}}>0$.
Then, $k$ runs from $1$ to $N_{s}$, where $1\leq N_{s}\leq2D-1$.
The corresponding critical times are $t_{k,m}^{c}=\frac{\left(2m-1\right)\pi}{2\omega_{q_{k}^{c},\delta}}$.
As a representative example, we plot in Fig.~\hyperref[fig:T-f]{\ref{fig:T-f}(b)}
$f(t)$ for $\delta_{0}=-\delta=0.5$ and $L=40\,000$. For these
parameters, we have that $N_{s}=3$ with $t_{1,1}^{c}\approx0.380\pi$,
$t_{1,2}^{c}\approx0.954\pi$, $t_{1,3}^{c}\approx1.591\pi$, and
$t_{2,1}^{c}\approx1.141\pi$. The corresponding non-analyticities
are cusps. Evidently, these cusps become rounded for finite systems
(see, for instance, the inset of Fig.~\hyperref[fig:T-f]{\ref{fig:T-f}(b)}).
However, for the case $\nu=0$, non-analyticities occur even for finite
systems (see Appendix \ref{AP-A}). 

\begin{figure}
\begin{centering}
\includegraphics[clip,width=0.8\columnwidth]{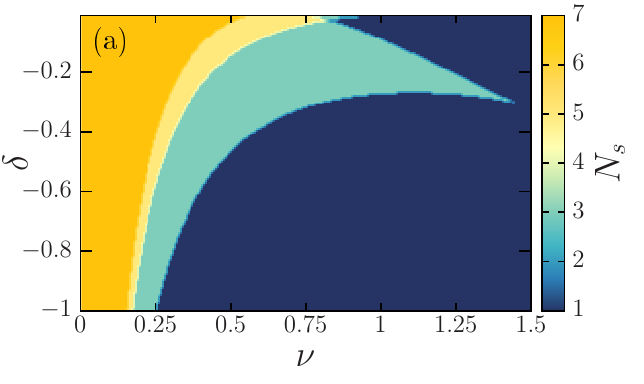}\\
\includegraphics[clip,width=0.8\columnwidth]{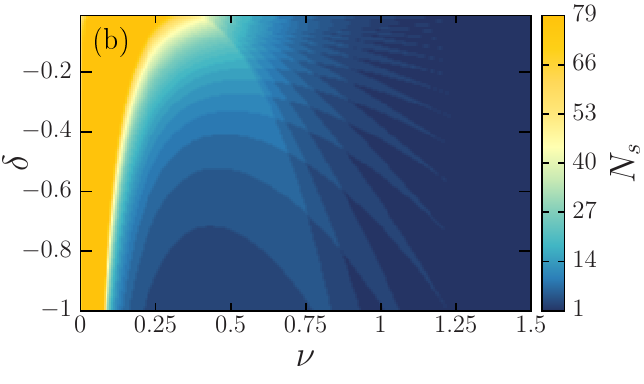}\\
\includegraphics[clip,scale=0.3]{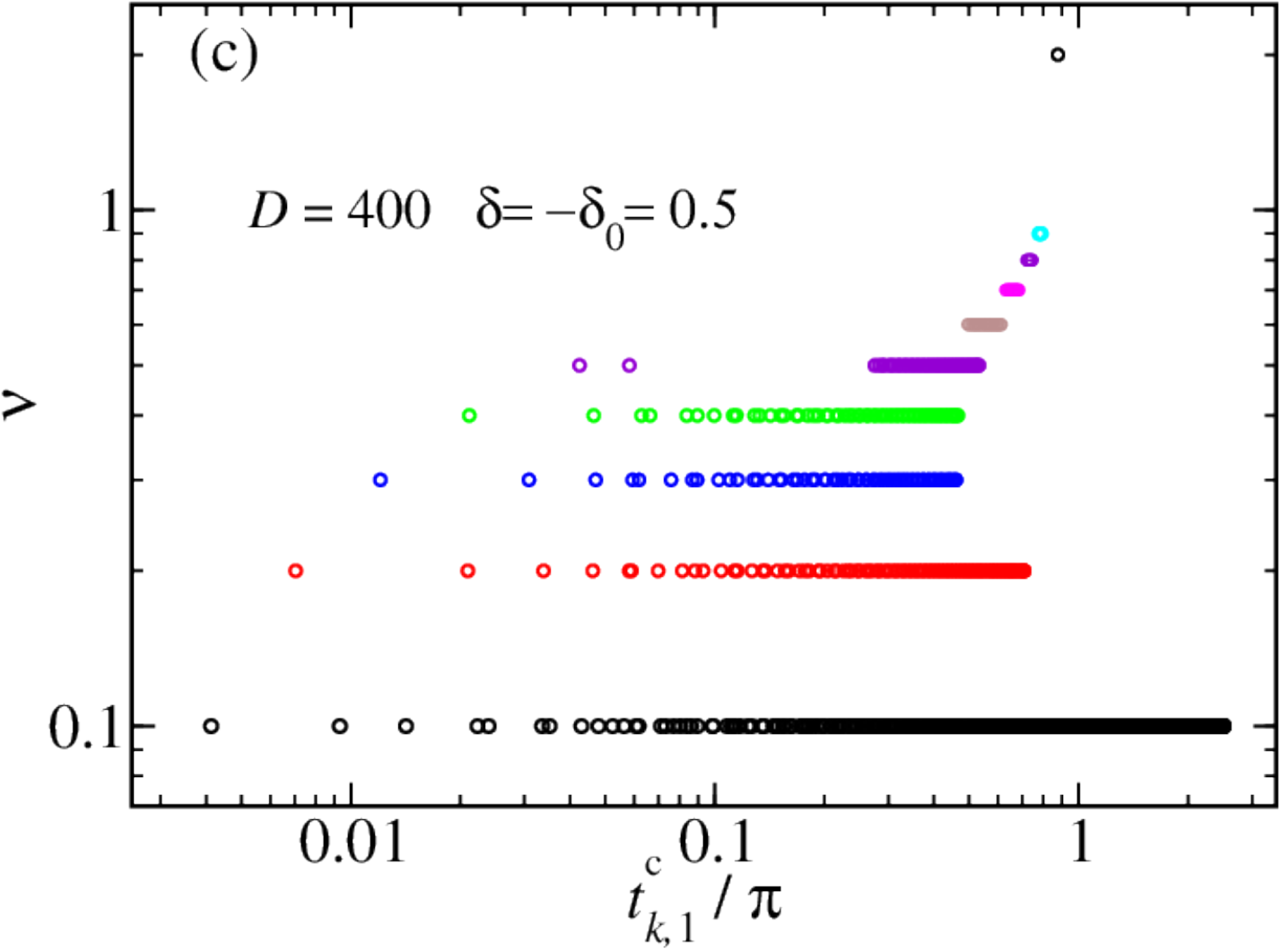}\\
\includegraphics[clip,scale=0.3]{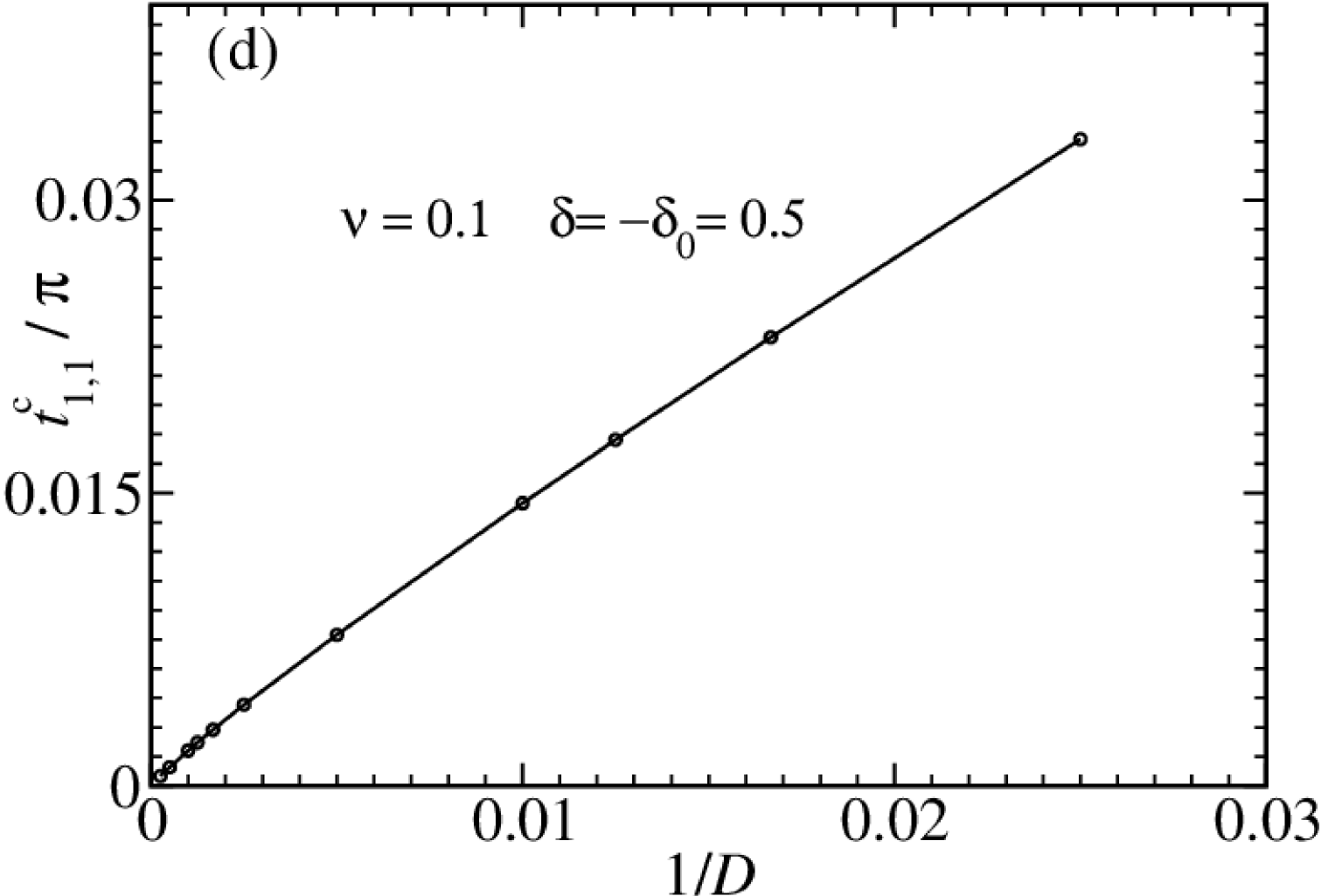}
\par\end{centering}
\caption{(a) The number of solutions of Eq.~(\ref{eq:T}), $N_{s}$, as a
function of $\nu$ and $\delta$ for fixed $\delta_{0}=1$ and (a)
$D=4$ and (b) $D=40$. (c) The critical times $t_{k,1}^{c}$ for
various values of $\nu$, $D=400$ and $\delta_{0}=-\delta=0.5$.
(d) The earliest critical instant $t_{1,1}^{c}$ vs. $1/D$ for $\nu=0.1$
and $\delta_{0}=-\delta=0.5$.\label{fig:N-tc} }
\end{figure}

As previously argued, the number of minima in $T_{q}^{2}$ is $D-1$
for sufficiently small $\nu$, yielding up to $N_{s}=2D-1$ solutions
of Eq.~(\ref{eq:T}) (critical time instants per accumulation line).
This number has to diminish when $\nu$ increases as $N_{s}=1$ for
$\nu\rightarrow\infty$. This is clearly demonstrated in Fig.~\hyperref[fig:T-f]{\ref{fig:T-f}(a)}
for $\nu=0.75$. Notice that, instead of only local minima, $T_{q}^{2}$
develops local maxima for larger values of $\nu$. This means that
the number of critical time instants $N_{s}$ per accumulation line
{[}solutions {]} is a non-monotonic function of the quench parameters
$\delta$ and $\delta_{0}$. This non-trivial behavior is demonstrated
in Figs.~\hyperref[fig:N-tc]{\ref{fig:N-tc}(a)} and \hyperref[fig:N-tc]{(b)}
where we plot $N_{s}$ as a function of $\nu$ and $\delta$ for fixed
$\delta_{0}=1$ and $D=4$ and $40$. Notice that $N_{s}$ always
change by $\pm2$ as these solutions always appear or disappear in
pairs. At the transition lines, two solutions degenerate. The resulting
non-analiticity is a smooth one as demonstrated for the case $D=2$. 

\begin{figure}[t]
\begin{centering}
\includegraphics[scale=0.35]{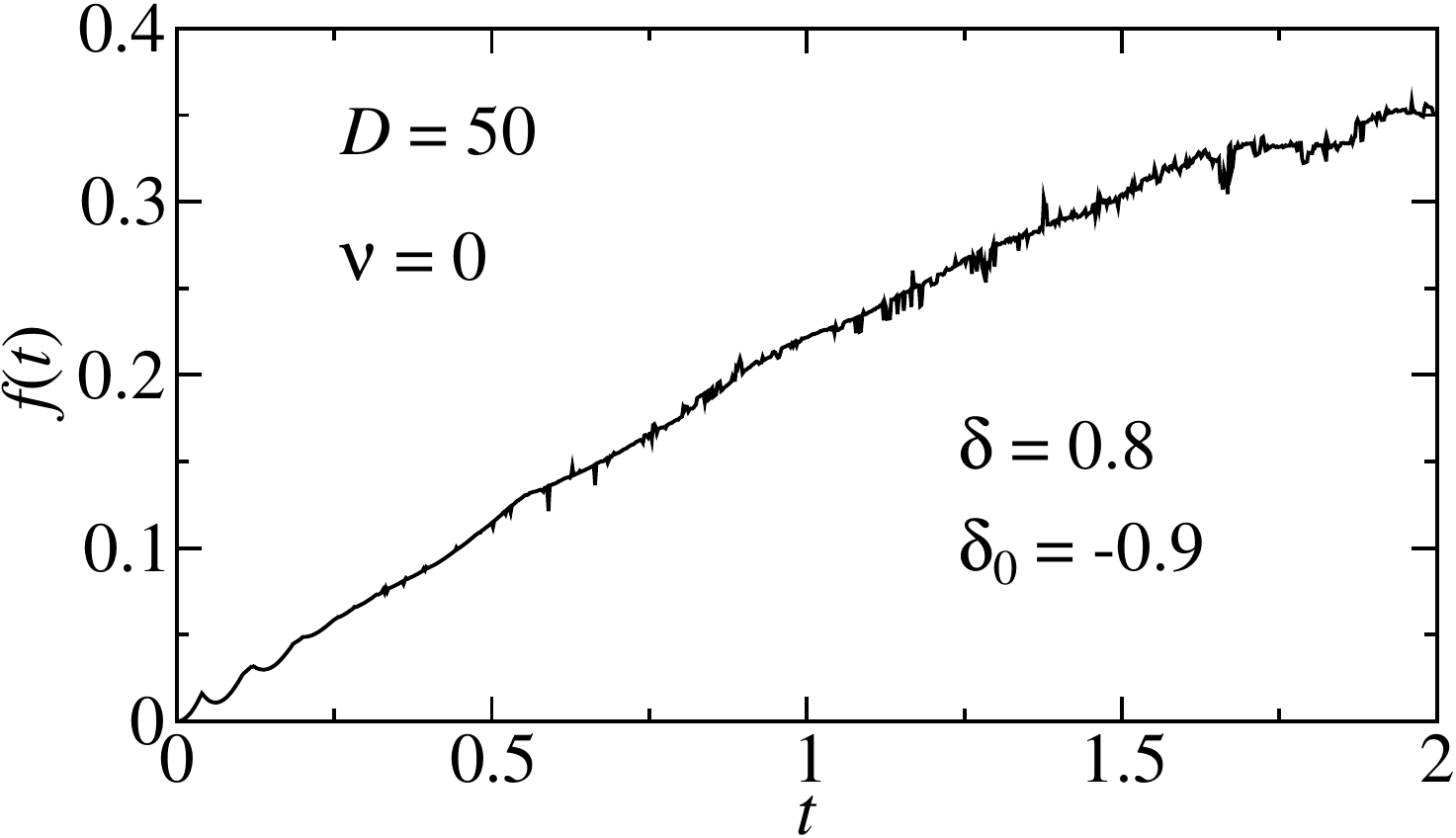}
\par\end{centering}
\caption{The dynamical free energy $f$ as a function of time $t$ in the case
$D=50$ and $\nu=0$ for the quantum quench from $\delta_{0}=-0.9$
to $\delta=0.8$. Many non-analyticities appear already at short time
scales.\label{fig:f-D50}}

\end{figure}

Having discussed the cases of large and small $\nu$, and small $D$,
we now discuss the interesting case of small $\nu$ and $D\gg1$.
As we have argued there can be $2D-1$ solutions of Eq.~(\ref{eq:T}).
This means the existence of many critical time instants per accumulation
line. More interesting, it can be demonstrated that the largest critical
time instant is of order unity and the smallest one is of order $D^{-1}$
{[}see Fig.~\hyperref[fig:N-tc]{\ref{fig:N-tc}(d)}{]}. As shown
in Fig.~\hyperref[fig:N-tc]{\ref{fig:N-tc}(c)}, these time instants
are somewhat evenly distributed in the interval $\left[\sim D^{-1},\sim1\right]$
(see more details in Appendix \ref{AP-A}). Intriguingly, this means
that for large values of $D$ the dynamic free energy $f(t)$ will
present a large number of non-analyticities in time. This is not only
because the number of critical time instants is of order $D$ per
accumulation line. As many of those instants happen at $t^{c}\sim D^{-1}$,
they ``reappear'' yet at short time-scales in the other accumulation
lines. As a result, $f(t)$ has \emph{non-analyticities at almost
all times} if the quantum quench crosses the transition, $D$ is sufficiently
large, and $\nu$ is sufficiently small (see Fig.~\ref{fig:f-D50}).

\section{Further discussions and conclusions\label{sec:CONCLUSION}}

We studied the dynamic free energy $f(t)$ of a free fermion chain
with long-range hopping couplings, which is described by Eq.~(\ref{eq:H-free-fermion-1}),
focusing on its non-analyticities and the associated Yang-Lee-Fisher
zeros. 

For effective short-range hoppings (small $D$ or large $\nu$) the
YLF zeros cross the real-time axis only in a few instants per accumulation
line. In contrast, when the hoppings are sufficiently long ranged
(large $D$ and small $\nu$), the number of times the YLF zeros cross
the real-time axis increases with $D$ and are more or less evenly
spread in the short time interval $0<Jt\lesssim1$, where $J$ is
the microscopic energy scale. 

We point out that these many non-analyticities are different from
other cases studied in the literature, where the YLF zeros accumulate
in an area on the complex-time plane. This is the case for the Kitaev
honeycomb model~\citep{schmitt-kehrein-prb15} and for disordered
systems exhibiting dynamical Griffiths singularities~\citep{NetoRafaelXavierPRBL2022}.
In the thermodynamic limit, the infinitely many zeros crossing the
real-time axis yield to non-analyticities only at the edges of those
distributions of zeros. Here, for the model Hamiltonian (\ref{eq:H-free-fermion-1}),
the zeros do not become continuously distributed over an area on the
complex-time plane. They remain distributed in lines that cross the
real-time axis in many different time instants. Evidently, when the
distance between these singularities increases beyond numerical or
experimental resolution, they will appear as a smooth function of
time, resembling the case of continuously distributed zeros over a
time window. 

We emphasize that the singularities are prominent only in sufficiently
large systems (rigorously, only in the thermodynamic limit), especially
when $D$ is large and $\nu$ small. Therefore, the observation of
these many singularities in the current cold-atom platform, where
the system size is not too large, may be a challenging task. Perhaps,
the best way to circumvent this obstacle is to consider model with
anti-periodic boundary condition, $D=L/4$, and $\nu=0$ (see Appendix
\ref{AP-A}). For this situation, the YLF zeros lie on the real-time
axis even for finite systems. We note that anti-periodic boundary
conditions can be realized by considering the one dimensional chain
with periodic boundary condition with a magnetic field passing through
the ring. For a particular choice of the flux magnetic, it is possible
to map this model to one with the anti-periodic boundary condition
(see, for instance, Refs.~\citealp{Twist1-PhysRevLett.7.46,Twist2-PhysRev.133.A171,twist3-PhysRevB.44.9562}).\textcolor{red}{{} }

To the best of our knowledge, long-range interaction effects in the
context of DQPTs have only been studied for the transverse-Field Ising
chain~\citep{heyl-trappexp,DQPTlograngisingPhysRevB.96.104436,DynamicLongRangePRB2017,DQPTlongrangePhysRevE.96.062118,PRLHeylLongRangDynamic}.
Although the model studied here is different, the present work may
shed light on what happens in other models. For instance, in the transverse-field
Ising model anomalous cusps (associated with the emergence of new
cusps) in the dynamic free energy were reported when $\nu\lesssim2.2$,
at least for some quench parameters~\citep{DynamicLongRangePRB2017}.
These cusps were denominated as anomalous simply because they are
not equally spaced in time. As we have explicitly shown, new cusps
not evenly separated in time appear for sufficiently long-range {hopping}
(small $\nu$) in a non-trivial fashion (see Fig.~\ref{fig:N-tc})
as predicted by Eq.~(\ref{eq:T}). It is then desirable to understand
Eq.~(\ref{eq:T}) in a more fundamental way and/or generalize it
to other systems, in particular, to non-integrable ones. To this end,
we recast Eq.~(\ref{eq:T}) is terms of general quantities and find
that it is equivalent to $\delta_{0}\omega_{q,\delta}^{2}+\delta\omega_{q,\delta_{0}}^{2}=0$.
Thus, in the lack of a better analogy, the number of YLF zeros (or
cusps) equals the number of Fermi point pairs of this ``weighted dispersion''
with zero ``chemical potential''. While this is a simple fact for
the model we studied, it would be desirable to verify it to other
models. For the conventional nearest-neighbor transvere-field Ising
chain, the analogous relation can be obtained by recasting the results
of Ref.~\citealp{HeylPRLseminalDynamic}: it is simply $\omega_{q,g}^{2}+\omega_{q,g_{0}}^{2}=\left(g-g_{0}\right)^{2}$,
where the dispersion relation is $\omega_{q,g}=\sqrt{\left(g-\cos q\right)^{2}+\sin^{2}q}$
and $g=h/J$ is the ratio between the transverse field and the ferromagnetic
coupling. Again, one needs to find the Fermi points of a weighted
dispersion with chemical potential $\left(g-g_{0}\right)^{2}$. We
emphasize that, in both models, the YLF zeros are determined uniquely
by the knowledge of the dispersion relation and of the pre- and post-quench
parameters. It certainly desirable to verify whether this remains
true for other models.

Finally, we mention that smaller the value of $\nu$, harder is the
detection of the non-analyticities numerically. In particular, the
cusps become rounded if the system size is not sufficiently large
{[}see Fig.~\ref{fig:T-f}{]} precluding its detection with exact
diagonalization. On the other hand, powerful numerical techniques
such as the tDMRG or the MPS use, typically, a time step $\Delta t\sim0.01/J$
to evolve the initial state. Our results indicate that such time step
is not sufficiently small to detect the non-analyticities that appear
already at short time scales when $1/(DJ)<0.01/J$ (or $D\gtrsim100).$
\begin{acknowledgments}
This research was supported by the Brazilian agencies FAPEMIG, CNPq,
and FAPESP. J.A.H. thanks IIT Madras for a visiting position under
the IoE program which facilitated the completion of this research
work.
\end{acknowledgments}

\appendix

\section{\label{AP-A}The case $\nu=0$ }

In this appendix, we consider the special case that $\nu=0$, where
Eqs.~(\ref{eq:C}) and (\ref{eq:S}) become
\begin{equation}
C_{q}=\frac{\sin\left(Dq\right)\cos\left(Dq\right)}{\sin q}\mbox{ and }S_{q}=\frac{\sin^{2}\left(Dq\right)}{\sin q}.\label{eq:ApCS}
\end{equation}

\subsection{Critical time instants}

We need to solve Eq.~(\ref{eq:T}) with the care of having $\omega_{q}\ne0$.
Thus, we need to solve 

\begin{equation}
\cos^{2}\left(Dq^{c}\right)+\delta_{0}\delta\sin^{2}\left(Dq^{c}\right)=0.\label{eq:ApCS2}
\end{equation}
 As we are interested in solutions in the interval $q\in[0,\frac{\pi}{2}]$,
then, 
\begin{equation}
q_{k}^{c}=\frac{1}{D}\left((k-1)\pi+\arcsin\left(\frac{1}{\sqrt{1-\delta\delta_{0}}}\right)\right),\ k=1,...,\frac{D}{2}.\label{eq:qc}
\end{equation}

As we already mentioned, in the Sec. \ref{sec:Results}, to solve
Eq. (\ref{eq:ApCS2}) we need that $D\ll L,$ otherwise, there are
not enough q's to satisfy this equation. Once we determine critical
values of $q$ that satisfy Eq. (\ref{eq:ApCS2}) we obtain the critical
times $t_{i,n}^{c}=\frac{(2n-1)\pi}{2\omega_{q_{i}^{c}}(\delta)}$,
$n=1,2,...$, which are given by

\begin{equation}
t_{k,m}^{c}=\left(m-\frac{1}{2}\right)\pi\frac{1-\delta\delta_{0}}{\sqrt{\delta(\delta-\delta_{0})}}\sin\left(q_{k}^{c}\right).\label{eq:tc}
\end{equation}

\subsubsection{The limit $D\gg1$}

In this limit, the first critical instants ($k\ll D$) of each accumulation
line $m$ become 
\begin{equation}
t_{k,m}^{c}\approx\left(m-\frac{1}{2}\right)\pi\frac{1-\delta\delta_{0}}{\sqrt{\delta(\delta-\delta_{0})}}\left(\frac{(k-1)\pi+\arcsin\left(\frac{1}{\sqrt{1-\delta\delta_{0}}}\right)}{D}\right).
\end{equation}
 Thus, they vanish $\sim D^{-1}$. 

\subsubsection{The case $D=L/4$ and $\delta\delta_{0}=-1$}

\begin{figure}[tp]
\begin{centering}
\includegraphics[clip,scale=0.35]{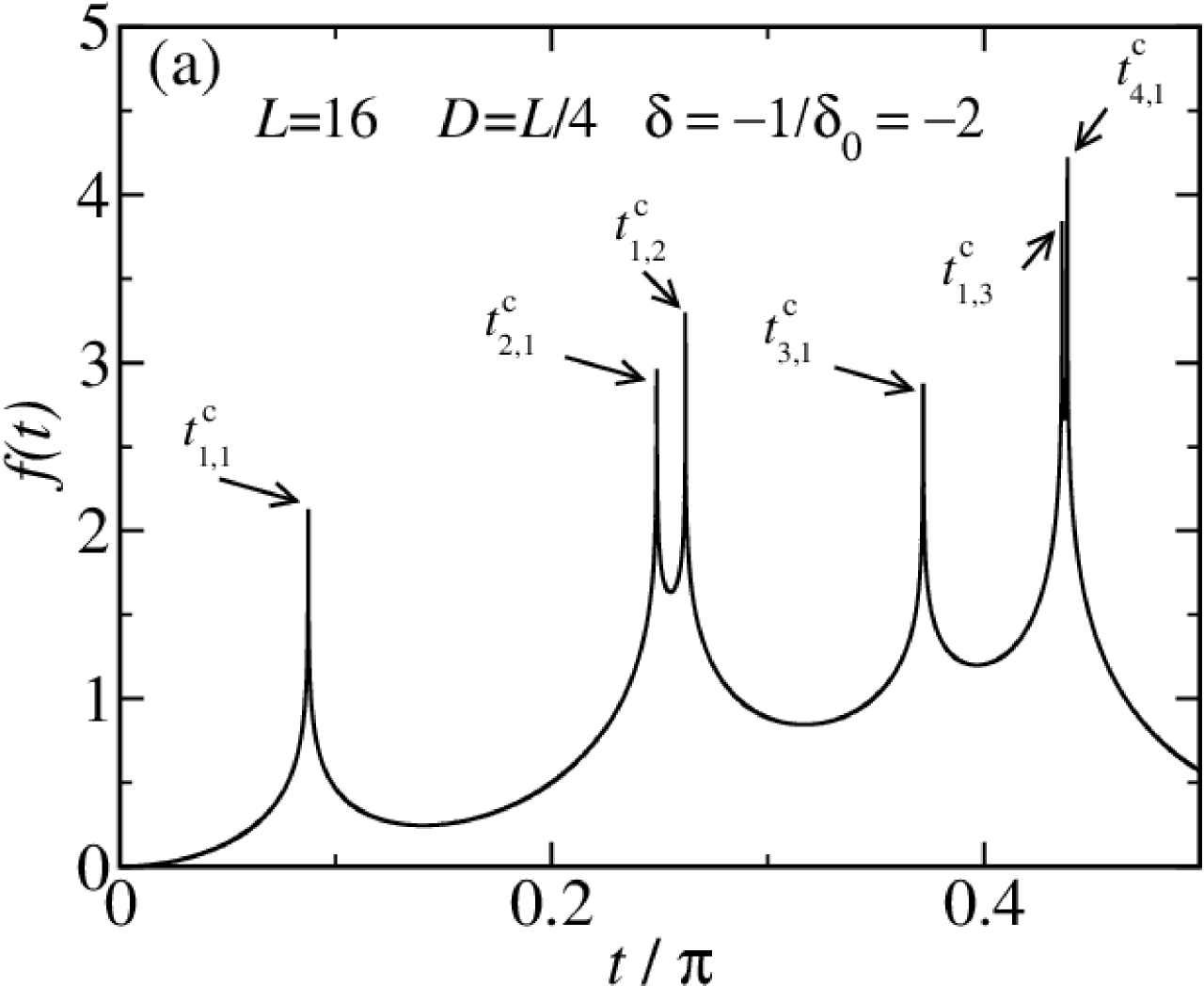}
\par\end{centering}
\begin{centering}
\includegraphics[clip,scale=0.34]{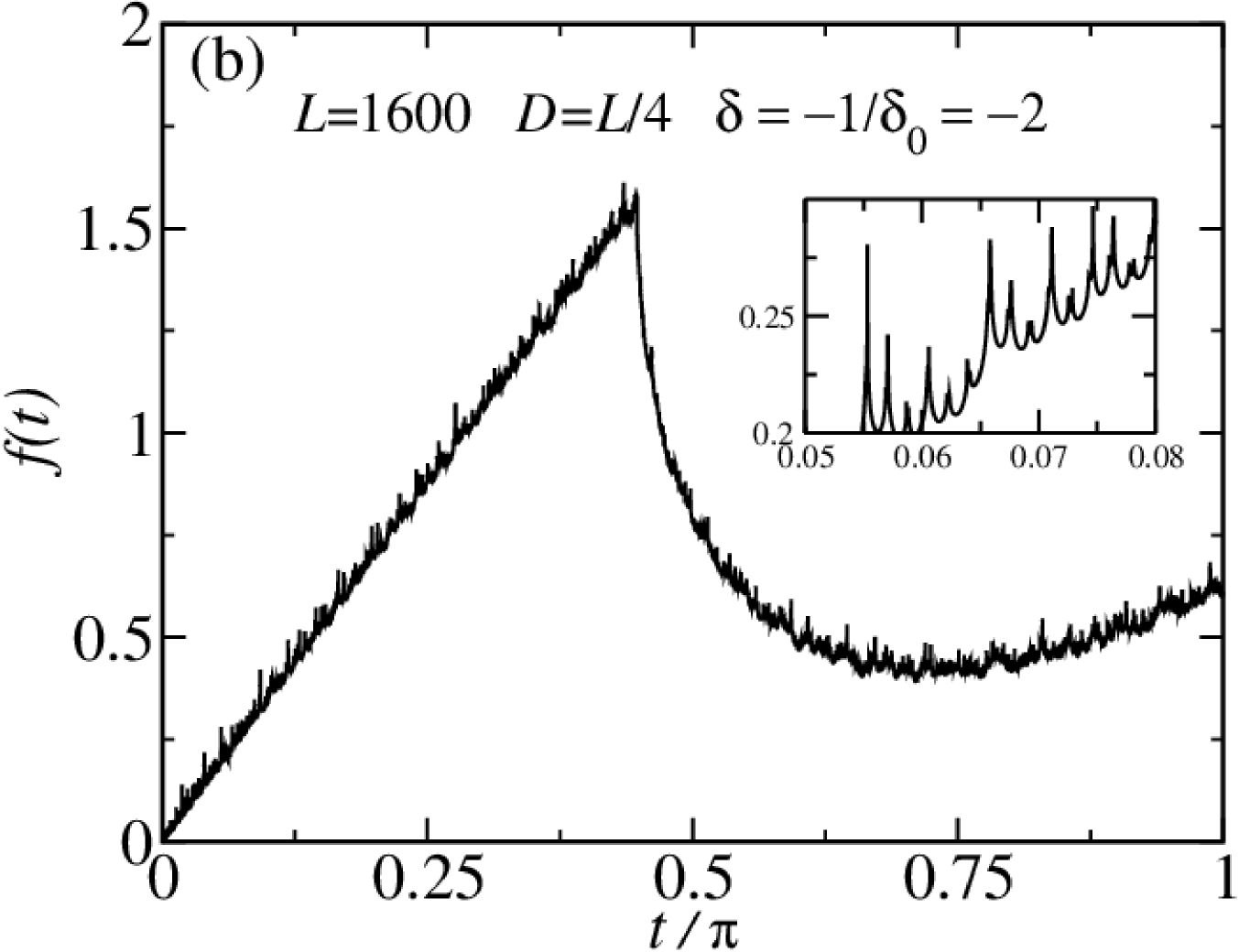}
\par\end{centering}
\caption{\label{figs2} (a) The dynamic free energy $f(t)$ vs. $t/\pi$ for
the case $\nu=0$, $D=L/4$, $\delta=-\frac{1}{\delta_{0}}=-2$. (a)
For a system size $L=16$. The arrows indicate the critical time positions
given by Eq.~(\ref{eq:tcnu0}). (b) The same as (a) but $L=1\,600$.
Inset: shows a zoom of the region close to $t=0.06\pi$. }
\end{figure}

When $\delta\delta_{0}=-1$, Eq.~(\ref{eq:qc}) becomes $q_{k}^{c}=\frac{\pi}{2D}\left(2k-\frac{3}{2}\right)$.
The Fourier wavevectors in (\ref{eq:Fourier}) are $q_{n}=\frac{2\pi}{L}\left(n-\frac{\phi}{2}\right)$.
Thus, interestingly, when $D=L/4$ and the anti-periodic boundary
condition is considered ($\phi=1$), all critical wavevectors $q_{k}^{c}$
exist even for finite systems (evidently, $L$ is a multiple of $4$).
The associated critical instants are 

\begin{equation}
t_{k,m}^{c}=\left(2m-1\right)\pi\frac{1}{\sqrt{1+\delta^{2}}}\sin\left(q_{k}^{c}\right).\label{eq:tcnu0}
\end{equation}
 Notice also that, because the zeros of $Z$ are on the real-time
axis even for finite systems, the dynamic free energy diverges at
$t_{m,k}^{c}$. Similar non-analyticities at finite systems were observed
in other models~\citep{DynamSirkerPRB2014,KarraschPhysRevB.87.195104,NetoRafaelXavierunpublish}.
We illustrate this peculiar behavior of the $f(t)$ in Fig.~\ref{figs2}
for a quench where $\delta=-\frac{1}{\delta_{0}}=2$ and $L=16$ and
$L=1\,600$. The peaks are finite due to the finite time step we used
($\sim10^{-4}$). Evidently, $f\left(t\right)$ becomes analytic in
the thermodynamic limit as there will be a continuous distribution
of YLF zeros over the real-time axis.

\subsection{Ground state energy for $D=L/4$ }

\begin{figure}[t]
\begin{centering}
\includegraphics[clip,scale=0.35]{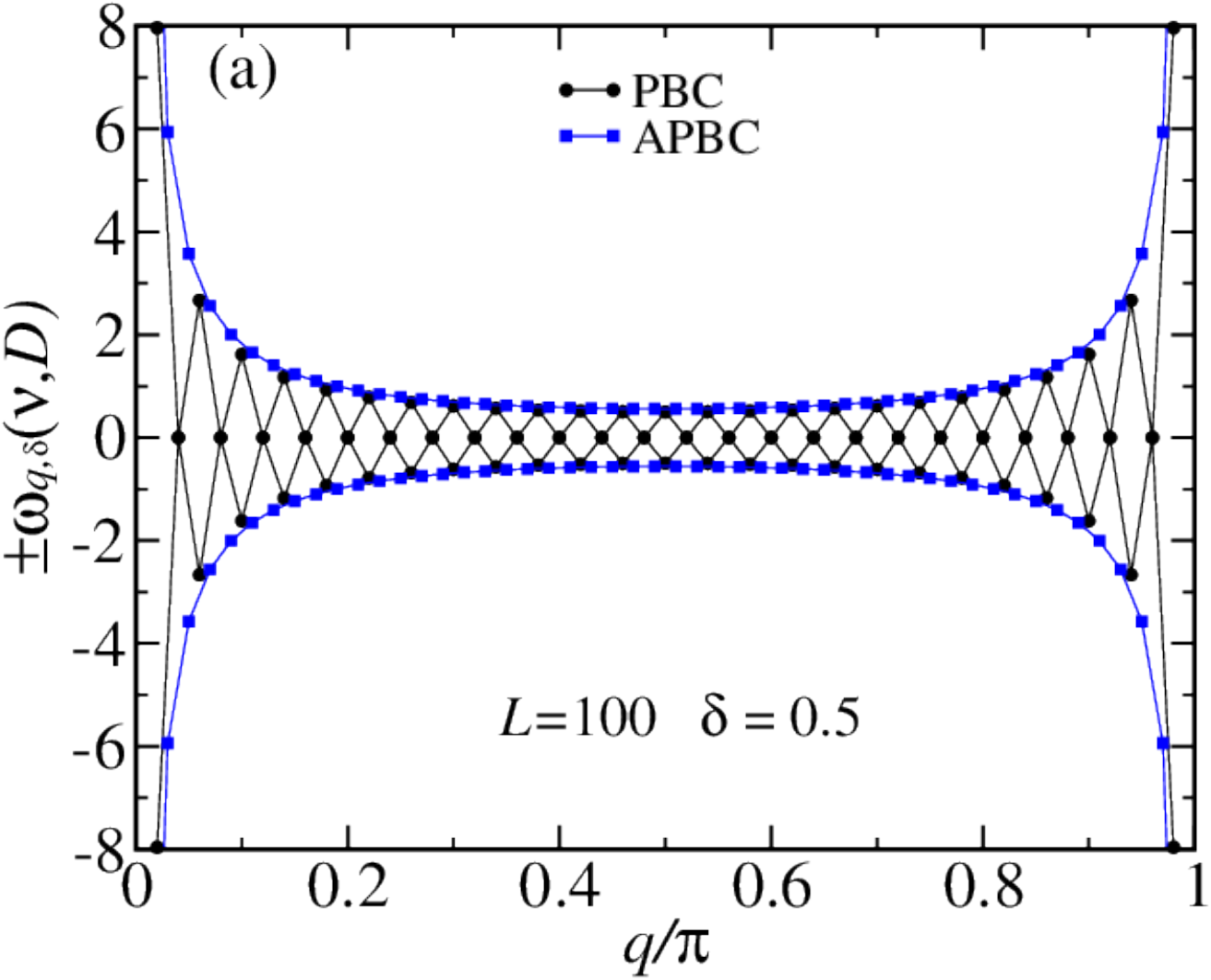}
\par\end{centering}
\begin{centering}
\includegraphics[clip,scale=0.35]{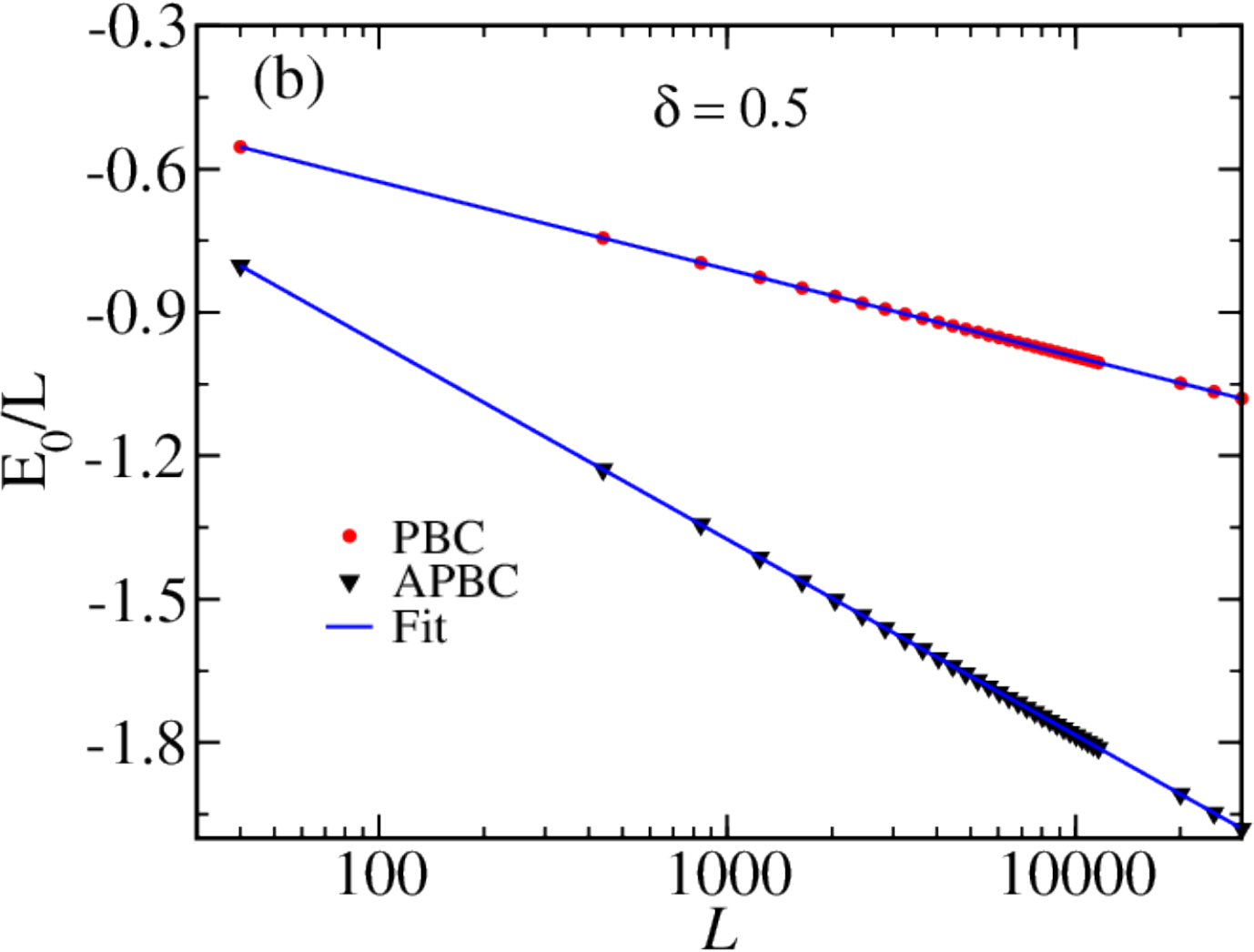}
\par\end{centering}
\caption{(a) The dispersion relation $\pm\omega_{q,\delta}(\nu,D)$ vs. $q$
for systems under PBC and APBC (see legend) with $L=100$, $\nu=0$,
$\delta=0.5$ and $D=L/4$. The symbols are the data obtained from
Eqs.~(\ref{eq:E0APBCnuExac}) and (\ref{eq:E0PBCnuExac}). The solid
lines connect the fitted points using Eqs. \ref{eq:E0APBCnu0a} and
\ref{eq:E0PBCnu0}. \label{fig:w-n0} }
\end{figure}

We now compute the ground state energy for systems with PBC ($\phi=0$)
and APBC ($\phi=1$), $\nu=0$, and $D=L/4$. The dispersion (\ref{eq:dispersion})
becomes\textcolor{red}{{} }
\begin{equation}
\omega_{q_{n},\delta}=\frac{\sqrt{\phi+\delta^{2}\left(1+2\left(1-\phi\right)\left(-1\right)^{n}\right)^{2}}}{2\sin q_{n}},\label{eq:w-n0}
\end{equation}
for $n=1,...,\frac{L}{2}$, except for $n=\frac{L}{2}$ and $\phi=0$.
Instead, in that case, $\omega_{\pi,\delta}=\frac{L}{4}$. Notice
that the system is gapless (gapful) for PBC (APBC) $\phi=0$ ($\phi=1$)
regardless of the value of the dimerization parameter $\delta$. A
similar situation appears in the topological insulators (TIs). However,
in the TIs the bulk is gapped under PBC and there are gapless boundary
states for OBC. In the present model, we have gapless states in the
bulk for the PBC case, and a gapped state for $\phi\ne0$. In Fig.~\hyperref[fig:w-n0]{\ref{fig:w-n0}(a)},
we illustrate the dispersion relation Eq.~(\ref{eq:w-n0}) for $L=100$
and $\delta=0.5$ for the model with PBC and APBC. It is interesting
to note that, in the thermodynamic limit, the system with PBC has
two degenerate flat bands.

The ground state energy $E_{0}^{\text{APBC}}(\delta,\nu=0)$ for the
system with APBC is 
\begin{equation}
E_{0}^{\text{APBC}}(\delta,0)=-\frac{\sqrt{1+\delta^{2}}}{2}\sum_{n=1}^{\frac{L}{2}}\frac{1}{\sin q_{n}}.\label{eq:E0APBCnuExac}
\end{equation}

We can replace a sum by a integral by using the Euler-Maclaurin sum 

\begin{equation}
\sum_{m=0}^{n}F(a+kh)=\frac{1}{h}\int_{a}^{b}F(q)dq+\frac{1}{2}\left(F(b)+F(a)\right)+R,
\end{equation}
where $R$ is the residual term. So,

\begin{eqnarray}
E_{0}^{\text{APBC}}(\delta,0) & = & -L\frac{\sqrt{1+\delta^{2}}}{4\pi}\Bigl\{\int_{\pi/L}^{\pi-\pi/L}\frac{dq}{\sin q}+{\displaystyle \frac{2\pi}{L\sin(\pi/L)}}\nonumber \\
 &  & +\frac{{\displaystyle 2\pi R}}{L}\Bigr\},\nonumber \\
 & = & -L\frac{\sqrt{1+\delta^{2}}}{2\pi}\Bigl\{\ln\left[\frac{\cos(\pi/2L)}{\sin(\pi/2L)}\right]+{\displaystyle \frac{\pi}{L\sin(\pi/L)}}\nonumber \\
 &  & +\pi a_{1}\Bigr\},\label{eq:E0APBCnu0a}
\end{eqnarray}
where we use the fact that the residual term $R=La_{1}$. We were
not able to obtain the exact value of $a_{1}$. However, we obtain
that $a_{1}=0.08596$ by fitting the exact data Eq.~(\ref{eq:E0APBCnuExac})
with Eq.~(\ref{eq:E0APBCnu0a}) {[}see Fig.~\hyperref[fig:w-n0]{\ref{fig:w-n0}(b)}{]}.
We verify that $a_{1}$ does not depend on $\delta$. For large values
of $L$, the energy per site becomes $E_{0}^{\text{APBC}}(\delta,0)/L=-\frac{\sqrt{1+\delta^{2}}}{2\pi}\left[\ln\left(2L/\pi\right)+2+\pi a_{1}\right]$.
Note that the energy is not extensive. However, we can recover the
extensivity if we consider the volume of the system as $V=L\ln L$.

For periodic boundary conditions, the ground state energy for $L$
multiple of $4$ is
\begin{equation}
E_{0}^{\text{PBC}}(\delta,0)=-\frac{L}{4}-\left|\delta\right|\sum_{m=0}^{\frac{L}{4}-1}\frac{1}{\sin\left(2\pi/L(2m+1)\right)}.\label{eq:E0PBCnuExac}
\end{equation}
Similarly as the APBC case, we obtain
\begin{eqnarray}
E_{0}^{\text{PBC}}(\delta,0) & = & -\frac{L}{4}-L\frac{\left|\delta\right|}{2\pi}\Bigl\{\ln\left[\frac{\cos(\pi/L)}{\sin(\pi/L)}\right]+{\displaystyle \frac{2\pi}{L\sin(2\pi/L)}}\nonumber \\
 &  & +2\pi a_{2}\Bigr\},\label{eq:E0PBCnu0}
\end{eqnarray}
where $a_{2}=0.04297\approx a_{1}/2$ as expected since the residual
term depends on the interval and on $h$, which are basically the
same in the thermodynamic limit. In this case, for large values of
$L$, the energy per site becomes $E_{0}^{\text{PBC}}(\delta,0)/L=-\frac{1}{4}-\frac{\left|\delta\right|}{2\pi}\left[\ln\left(L/\pi\right)+2+2\pi a_{2}\right]$.
Note that $E_{0}^{\text{APBC}}-E_{0}^{\text{PBC}}\sim-\left(\sqrt{1+\delta^{2}}-\left|\delta\right|\right)L\ln(L$). 

\phantomsection\addcontentsline{toc}{section}{\refname}\bibliography{refs_rev4}

\begin{thebibliography}{50}%
\makeatletter
\providecommand \@ifxundefined [1]{%
 \@ifx{#1\undefined}
}%
\providecommand \@ifnum [1]{%
 \ifnum #1\expandafter \@firstoftwo
 \else \expandafter \@secondoftwo
 \fi
}%
\providecommand \@ifx [1]{%
 \ifx #1\expandafter \@firstoftwo
 \else \expandafter \@secondoftwo
 \fi
}%
\providecommand \natexlab [1]{#1}%
\providecommand \enquote  [1]{``#1''}%
\providecommand \bibnamefont  [1]{#1}%
\providecommand \bibfnamefont [1]{#1}%
\providecommand \citenamefont [1]{#1}%
\providecommand \href@noop [0]{\@secondoftwo}%
\providecommand \href [0]{\begingroup \@sanitize@url \@href}%
\providecommand \@href[1]{\@@startlink{#1}\@@href}%
\providecommand \@@href[1]{\endgroup#1\@@endlink}%
\providecommand \@sanitize@url [0]{\catcode `\\12\catcode `\$12\catcode
  `\&12\catcode `\#12\catcode `\^12\catcode `\_12\catcode `\%12\relax}%
\providecommand \@@startlink[1]{}%
\providecommand \@@endlink[0]{}%
\providecommand \url  [0]{\begingroup\@sanitize@url \@url }%
\providecommand \@url [1]{\endgroup\@href {#1}{\urlprefix }}%
\providecommand \urlprefix  [0]{URL }%
\providecommand \Eprint [0]{\href }%
\providecommand \doibase [0]{http://dx.doi.org/}%
\providecommand \selectlanguage [0]{\@gobble}%
\providecommand \bibinfo  [0]{\@secondoftwo}%
\providecommand \bibfield  [0]{\@secondoftwo}%
\providecommand \translation [1]{[#1]}%
\providecommand \BibitemOpen [0]{}%
\providecommand \bibitemStop [0]{}%
\providecommand \bibitemNoStop [0]{.\EOS\space}%
\providecommand \EOS [0]{\spacefactor3000\relax}%
\providecommand \BibitemShut  [1]{\csname bibitem#1\endcsname}%
\let\auto@bib@innerbib\@empty
%</preamble>
\bibitem [{\citenamefont {{S. Sachdev}}(2011)}]{booksachdev}%
  \BibitemOpen
  \bibfield  {author} {\bibinfo {author} {\bibnamefont {{S. Sachdev}}},\
  }\href@noop {} {\emph {\bibinfo {title} {Quantum {P}hase {T}ransitions}}},\
  \bibinfo {edition} {2nd}\ ed.\ (\bibinfo  {publisher} {Cambridge University
  Press},\ \bibinfo {year} {2011})\BibitemShut {NoStop}%
\bibitem [{\citenamefont {{H. E. Stanley}}(1987)}]{bookStanley}%
  \BibitemOpen
  \bibfield  {author} {\bibinfo {author} {\bibnamefont {{H. E. Stanley}}},\
  }\href@noop {} {\emph {\bibinfo {title} {Introduction to {P}hase
  {T}ransitions and {C}ritical {P}henomena}}}\ (\bibinfo  {publisher} {Oxford
  University Press},\ \bibinfo {year} {1987})\BibitemShut {NoStop}%
\bibitem [{\citenamefont {Sondhi}\ \emph {et~al.}(1997)\citenamefont {Sondhi},
  \citenamefont {Girvin}, \citenamefont {Carini},\ and\ \citenamefont
  {Shahar}}]{RevPT1997}%
  \BibitemOpen
  \bibfield  {author} {\bibinfo {author} {\bibfnamefont {S.~L.}\ \bibnamefont
  {Sondhi}}, \bibinfo {author} {\bibfnamefont {S.~M.}\ \bibnamefont {Girvin}},
  \bibinfo {author} {\bibfnamefont {J.~P.}\ \bibnamefont {Carini}}, \ and\
  \bibinfo {author} {\bibfnamefont {D.}~\bibnamefont {Shahar}},\ }\bibfield
  {title} {\enquote {\bibinfo {title} {Continuous quantum phase transitions},}\
  }\href {\doibase 10.1103/RevModPhys.69.315} {\bibfield  {journal} {\bibinfo
  {journal} {Rev. Mod. Phys.}\ }\textbf {\bibinfo {volume} {69}},\ \bibinfo
  {pages} {315--333} (\bibinfo {year} {1997})}\BibitemShut {NoStop}%
\bibitem [{\citenamefont {Yang}\ and\ \citenamefont {Lee}(1952)}]{YLzerosI}%
  \BibitemOpen
  \bibfield  {author} {\bibinfo {author} {\bibfnamefont {C.~N.}\ \bibnamefont
  {Yang}}\ and\ \bibinfo {author} {\bibfnamefont {T.~D.}\ \bibnamefont {Lee}},\
  }\bibfield  {title} {\enquote {\bibinfo {title} {Statistical theory of
  equations of state and phase transitions. i. theory of condensation},}\
  }\href {\doibase 10.1103/PhysRev.87.404} {\bibfield  {journal} {\bibinfo
  {journal} {Phys. Rev.}\ }\textbf {\bibinfo {volume} {87}},\ \bibinfo {pages}
  {404--409} (\bibinfo {year} {1952})}\BibitemShut {NoStop}%
\bibitem [{\citenamefont {Lee}\ and\ \citenamefont {Yang}(1952)}]{YLzerosII}%
  \BibitemOpen
  \bibfield  {author} {\bibinfo {author} {\bibfnamefont {T.~D.}\ \bibnamefont
  {Lee}}\ and\ \bibinfo {author} {\bibfnamefont {C.~N.}\ \bibnamefont {Yang}},\
  }\bibfield  {title} {\enquote {\bibinfo {title} {Statistical theory of
  equations of state and phase transitions. ii. lattice gas and ising model},}\
  }\href {\doibase 10.1103/PhysRev.87.410} {\bibfield  {journal} {\bibinfo
  {journal} {Phys. Rev.}\ }\textbf {\bibinfo {volume} {87}},\ \bibinfo {pages}
  {410--419} (\bibinfo {year} {1952})}\BibitemShut {NoStop}%
\bibitem [{\citenamefont {{M. E. Fisher}}(1965)}]{bookFisherYLzeros}%
  \BibitemOpen
  \bibfield  {author} {\bibinfo {author} {\bibnamefont {{M. E. Fisher}}},\
  }\href@noop {} {\emph {\bibinfo {title} {The Nature of Critical Points}}},\
  \bibinfo {edition} {({L}ectures in {T}heoretical {P}hysic. {V}ol. viic)}\
  ed.\ (\bibinfo  {publisher} {{W}. {E}. {B}rittin, {N}ew {Y}ork: {G}olden and
  {B}reach},\ \bibinfo {year} {1965})\BibitemShut {NoStop}%
\bibitem [{\citenamefont {Wei}\ and\ \citenamefont
  {Liu}(2012)}]{PRLYLFzeros2012}%
  \BibitemOpen
  \bibfield  {author} {\bibinfo {author} {\bibfnamefont {Bo-Bo}\ \bibnamefont
  {Wei}}\ and\ \bibinfo {author} {\bibfnamefont {Ren-Bao}\ \bibnamefont
  {Liu}},\ }\bibfield  {title} {\enquote {\bibinfo {title} {Lee-yang zeros and
  critical times in decoherence of a probe spin coupled to a bath},}\ }\href
  {\doibase 10.1103/PhysRevLett.109.185701} {\bibfield  {journal} {\bibinfo
  {journal} {Phys. Rev. Lett.}\ }\textbf {\bibinfo {volume} {109}},\ \bibinfo
  {pages} {185701} (\bibinfo {year} {2012})}\BibitemShut {NoStop}%
\bibitem [{\citenamefont {Peng}\ \emph {et~al.}(2015)\citenamefont {Peng},
  \citenamefont {Zhou}, \citenamefont {Wei}, \citenamefont {Cui}, \citenamefont
  {Du},\ and\ \citenamefont {Liu}}]{PRLYLFzeros2015}%
  \BibitemOpen
  \bibfield  {author} {\bibinfo {author} {\bibfnamefont {Xinhua}\ \bibnamefont
  {Peng}}, \bibinfo {author} {\bibfnamefont {Hui}\ \bibnamefont {Zhou}},
  \bibinfo {author} {\bibfnamefont {Bo-Bo}\ \bibnamefont {Wei}}, \bibinfo
  {author} {\bibfnamefont {Jiangyu}\ \bibnamefont {Cui}}, \bibinfo {author}
  {\bibfnamefont {Jiangfeng}\ \bibnamefont {Du}}, \ and\ \bibinfo {author}
  {\bibfnamefont {Ren-Bao}\ \bibnamefont {Liu}},\ }\bibfield  {title} {\enquote
  {\bibinfo {title} {Experimental observation of lee-yang zeros},}\ }\href
  {\doibase 10.1103/PhysRevLett.114.010601} {\bibfield  {journal} {\bibinfo
  {journal} {Phys. Rev. Lett.}\ }\textbf {\bibinfo {volume} {114}},\ \bibinfo
  {pages} {010601} (\bibinfo {year} {2015})}\BibitemShut {NoStop}%
\bibitem [{\citenamefont {{J. L. Cardy}}(1989)}]{Cardyboundary}%
  \BibitemOpen
  \bibfield  {author} {\bibinfo {author} {\bibnamefont {{J. L. Cardy}}},\
  }\bibfield  {title} {\enquote {\bibinfo {title} {Boundary conditions, fusion
  rules and the {V}erlinde formula},}\ }\href {\doibase
  10.1016/0550-3213(89)90521-X} {\bibfield  {journal} {\bibinfo  {journal}
  {Nucl. Phys. B}\ }\textbf {\bibinfo {volume} {324}},\ \bibinfo {pages} {581}
  (\bibinfo {year} {1989})}\BibitemShut {NoStop}%
\bibitem [{\citenamefont {{J. L. Cardy}}\ and\ \citenamefont {{D. C.
  Lewellen}}(1991)}]{cardyboundary2}%
  \BibitemOpen
  \bibfield  {author} {\bibinfo {author} {\bibnamefont {{J. L. Cardy}}}\ and\
  \bibinfo {author} {\bibnamefont {{D. C. Lewellen}}},\ }\bibfield  {title}
  {\enquote {\bibinfo {title} {Bulk and boundary operators in conformal field
  theory},}\ }\href {\doibase 10.1016/0370-2693(91)90828-E} {\bibfield
  {journal} {\bibinfo  {journal} {Phys. Lett. B}\ }\textbf {\bibinfo {volume}
  {259}},\ \bibinfo {pages} {274} (\bibinfo {year} {1991})}\BibitemShut
  {NoStop}%
\bibitem [{\citenamefont {LeClair}\ \emph {et~al.}(1995)\citenamefont
  {LeClair}, \citenamefont {Mussardo}, \citenamefont {Saleur},\ and\
  \citenamefont {Skorik}}]{LECLAIboundary}%
  \BibitemOpen
  \bibfield  {author} {\bibinfo {author} {\bibfnamefont {A.}~\bibnamefont
  {LeClair}}, \bibinfo {author} {\bibfnamefont {G.}~\bibnamefont {Mussardo}},
  \bibinfo {author} {\bibfnamefont {H.}~\bibnamefont {Saleur}}, \ and\ \bibinfo
  {author} {\bibfnamefont {S.}~\bibnamefont {Skorik}},\ }\bibfield  {title}
  {\enquote {\bibinfo {title} {Boundary energy and boundary states in
  integrable quantum field theories},}\ }\href {\doibase
  https://doi.org/10.1016/0550-3213(95)00435-U} {\bibfield  {journal} {\bibinfo
   {journal} {Nuclear Physics B}\ }\textbf {\bibinfo {volume} {453}},\ \bibinfo
  {pages} {581--618} (\bibinfo {year} {1995})}\BibitemShut {NoStop}%
\bibitem [{\citenamefont {Heyl}\ \emph {et~al.}(2013)\citenamefont {Heyl},
  \citenamefont {Polkovnikov},\ and\ \citenamefont
  {Kehrein}}]{HeylPRLseminalDynamic}%
  \BibitemOpen
  \bibfield  {author} {\bibinfo {author} {\bibfnamefont {M.}~\bibnamefont
  {Heyl}}, \bibinfo {author} {\bibfnamefont {A.}~\bibnamefont {Polkovnikov}}, \
  and\ \bibinfo {author} {\bibfnamefont {S.}~\bibnamefont {Kehrein}},\
  }\bibfield  {title} {\enquote {\bibinfo {title} {Dynamical quantum phase
  transitions in the transverse-field ising model},}\ }\href {\doibase
  10.1103/PhysRevLett.110.135704} {\bibfield  {journal} {\bibinfo  {journal}
  {Phys. Rev. Lett.}\ }\textbf {\bibinfo {volume} {110}},\ \bibinfo {pages}
  {135704} (\bibinfo {year} {2013})}\BibitemShut {NoStop}%
\bibitem [{\citenamefont {Andraschko}\ and\ \citenamefont
  {Sirker}(2014)}]{DynamSirkerPRB2014}%
  \BibitemOpen
  \bibfield  {author} {\bibinfo {author} {\bibfnamefont {F.}~\bibnamefont
  {Andraschko}}\ and\ \bibinfo {author} {\bibfnamefont {J.}~\bibnamefont
  {Sirker}},\ }\bibfield  {title} {\enquote {\bibinfo {title} {Dynamical
  quantum phase transitions and the loschmidt echo: A transfer matrix
  approach},}\ }\href {\doibase 10.1103/PhysRevB.89.125120} {\bibfield
  {journal} {\bibinfo  {journal} {Phys. Rev. B}\ }\textbf {\bibinfo {volume}
  {89}},\ \bibinfo {pages} {125120} (\bibinfo {year} {2014})}\BibitemShut
  {NoStop}%
\bibitem [{\citenamefont {Vajna}\ and\ \citenamefont
  {D\'ora}(2014)}]{DynamicVajnaPRB2014}%
  \BibitemOpen
  \bibfield  {author} {\bibinfo {author} {\bibfnamefont {Szabolcs}\
  \bibnamefont {Vajna}}\ and\ \bibinfo {author} {\bibfnamefont {Bal\'azs}\
  \bibnamefont {D\'ora}},\ }\bibfield  {title} {\enquote {\bibinfo {title}
  {Disentangling dynamical phase transitions from equilibrium phase
  transitions},}\ }\href {\doibase 10.1103/PhysRevB.89.161105} {\bibfield
  {journal} {\bibinfo  {journal} {Phys. Rev. B}\ }\textbf {\bibinfo {volume}
  {89}},\ \bibinfo {pages} {161105} (\bibinfo {year} {2014})}\BibitemShut
  {NoStop}%
\bibitem [{\citenamefont {Karrasch}\ and\ \citenamefont
  {Schuricht}(2013)}]{KarraschPhysRevB.87.195104}%
  \BibitemOpen
  \bibfield  {author} {\bibinfo {author} {\bibfnamefont {C.}~\bibnamefont
  {Karrasch}}\ and\ \bibinfo {author} {\bibfnamefont {D.}~\bibnamefont
  {Schuricht}},\ }\bibfield  {title} {\enquote {\bibinfo {title} {Dynamical
  phase transitions after quenches in nonintegrable models},}\ }\href {\doibase
  10.1103/PhysRevB.87.195104} {\bibfield  {journal} {\bibinfo  {journal} {Phys.
  Rev. B}\ }\textbf {\bibinfo {volume} {87}},\ \bibinfo {pages} {195104}
  (\bibinfo {year} {2013})}\BibitemShut {NoStop}%
\bibitem [{\citenamefont {Canovi}\ \emph {et~al.}(2014)\citenamefont {Canovi},
  \citenamefont {Werner},\ and\ \citenamefont
  {Eckstein}}]{FirtOderDynanimcPRL}%
  \BibitemOpen
  \bibfield  {author} {\bibinfo {author} {\bibfnamefont {Elena}\ \bibnamefont
  {Canovi}}, \bibinfo {author} {\bibfnamefont {Philipp}\ \bibnamefont
  {Werner}}, \ and\ \bibinfo {author} {\bibfnamefont {Martin}\ \bibnamefont
  {Eckstein}},\ }\bibfield  {title} {\enquote {\bibinfo {title} {First-order
  dynamical phase transitions},}\ }\href {\doibase
  10.1103/PhysRevLett.113.265702} {\bibfield  {journal} {\bibinfo  {journal}
  {Phys. Rev. Lett.}\ }\textbf {\bibinfo {volume} {113}},\ \bibinfo {pages}
  {265702} (\bibinfo {year} {2014})}\BibitemShut {NoStop}%
\bibitem [{\citenamefont {Vajna}\ and\ \citenamefont
  {D\'ora}(2015)}]{PRBVajnaDynamic}%
  \BibitemOpen
  \bibfield  {author} {\bibinfo {author} {\bibfnamefont {Szabolcs}\
  \bibnamefont {Vajna}}\ and\ \bibinfo {author} {\bibfnamefont {Bal\'azs}\
  \bibnamefont {D\'ora}},\ }\bibfield  {title} {\enquote {\bibinfo {title}
  {Topological classification of dynamical phase transitions},}\ }\href
  {\doibase 10.1103/PhysRevB.91.155127} {\bibfield  {journal} {\bibinfo
  {journal} {Phys. Rev. B}\ }\textbf {\bibinfo {volume} {91}},\ \bibinfo
  {pages} {155127} (\bibinfo {year} {2015})}\BibitemShut {NoStop}%
\bibitem [{\citenamefont {Halimeh}\ and\ \citenamefont
  {Zauner-Stauber}(2017)}]{DynamicLongRangePRB2017}%
  \BibitemOpen
  \bibfield  {author} {\bibinfo {author} {\bibfnamefont {Jad~C.}\ \bibnamefont
  {Halimeh}}\ and\ \bibinfo {author} {\bibfnamefont {Valentin}\ \bibnamefont
  {Zauner-Stauber}},\ }\bibfield  {title} {\enquote {\bibinfo {title}
  {Dynamical phase diagram of quantum spin chains with long-range
  interactions},}\ }\href {\doibase 10.1103/PhysRevB.96.134427} {\bibfield
  {journal} {\bibinfo  {journal} {Phys. Rev. B}\ }\textbf {\bibinfo {volume}
  {96}},\ \bibinfo {pages} {134427} (\bibinfo {year} {2017})}\BibitemShut
  {NoStop}%
\bibitem [{\citenamefont {\ifmmode \check{Z}\else
  \v{Z}\fi{}unkovi\ifmmode~\check{c}\else \v{c}\fi{}}\ \emph
  {et~al.}(2018)\citenamefont {\ifmmode \check{Z}\else
  \v{Z}\fi{}unkovi\ifmmode~\check{c}\else \v{c}\fi{}}, \citenamefont {Heyl},
  \citenamefont {Knap},\ and\ \citenamefont {Silva}}]{PRLHeylLongRangDynamic}%
  \BibitemOpen
  \bibfield  {author} {\bibinfo {author} {\bibfnamefont {Bojan}\ \bibnamefont
  {\ifmmode \check{Z}\else \v{Z}\fi{}unkovi\ifmmode~\check{c}\else
  \v{c}\fi{}}}, \bibinfo {author} {\bibfnamefont {Markus}\ \bibnamefont
  {Heyl}}, \bibinfo {author} {\bibfnamefont {Michael}\ \bibnamefont {Knap}}, \
  and\ \bibinfo {author} {\bibfnamefont {Alessandro}\ \bibnamefont {Silva}},\
  }\bibfield  {title} {\enquote {\bibinfo {title} {Dynamical quantum phase
  transitions in spin chains with long-range interactions: Merging different
  concepts of nonequilibrium criticality},}\ }\href {\doibase
  10.1103/PhysRevLett.120.130601} {\bibfield  {journal} {\bibinfo  {journal}
  {Phys. Rev. Lett.}\ }\textbf {\bibinfo {volume} {120}},\ \bibinfo {pages}
  {130601} (\bibinfo {year} {2018})}\BibitemShut {NoStop}%
\bibitem [{\citenamefont {Fl\"aschner}\ \emph {et~al.}(2018)\citenamefont
  {Fl\"aschner}, \citenamefont {Vogel}, \citenamefont {Tarnowski},
  \citenamefont {Rem}, \citenamefont {L\"uhmann}, \citenamefont {Budich},
  \citenamefont {Mathey}, \citenamefont {Sengstock},\ and\ \citenamefont
  {Weitenberg}}]{DQPTexpnatur2018}%
  \BibitemOpen
  \bibfield  {author} {\bibinfo {author} {\bibfnamefont {N.}~\bibnamefont
  {Fl\"aschner}}, \bibinfo {author} {\bibfnamefont {D.}~\bibnamefont {Vogel}},
  \bibinfo {author} {\bibfnamefont {M.}~\bibnamefont {Tarnowski}}, \bibinfo
  {author} {\bibfnamefont {B.~S.}\ \bibnamefont {Rem}}, \bibinfo {author}
  {\bibfnamefont {D.-S.}\ \bibnamefont {L\"uhmann}}, \bibinfo {author}
  {\bibfnamefont {J.~C.}\ \bibnamefont {Budich}}, \bibinfo {author}
  {\bibfnamefont {L.}~\bibnamefont {Mathey}}, \bibinfo {author} {\bibfnamefont
  {K.}~\bibnamefont {Sengstock}}, \ and\ \bibinfo {author} {\bibfnamefont
  {C.}~\bibnamefont {Weitenberg}},\ }\bibfield  {title} {\enquote {\bibinfo
  {title} {Quasiparticle engineering and entanglement propagation in a quantum
  many-body system},}\ }\href {\doibase 10.1038/s41567-017-0013-8} {\bibfield
  {journal} {\bibinfo  {journal} {Nature Physcs}\ }\textbf {\bibinfo {volume}
  {14}},\ \bibinfo {pages} {265} (\bibinfo {year} {2018})}\BibitemShut
  {NoStop}%
\bibitem [{\citenamefont {Jafari}\ \emph {et~al.}(2019)\citenamefont {Jafari},
  \citenamefont {Johannesson}, \citenamefont {Langari},\ and\ \citenamefont
  {Martin-Delgado}}]{dynamicDelgadoPhysRevB.99.054302}%
  \BibitemOpen
  \bibfield  {author} {\bibinfo {author} {\bibfnamefont {R.}~\bibnamefont
  {Jafari}}, \bibinfo {author} {\bibfnamefont {Henrik}\ \bibnamefont
  {Johannesson}}, \bibinfo {author} {\bibfnamefont {A.}~\bibnamefont
  {Langari}}, \ and\ \bibinfo {author} {\bibfnamefont {M.~A.}\ \bibnamefont
  {Martin-Delgado}},\ }\bibfield  {title} {\enquote {\bibinfo {title} {Quench
  dynamics and zero-energy modes: The case of the creutz model},}\ }\href
  {\doibase 10.1103/PhysRevB.99.054302} {\bibfield  {journal} {\bibinfo
  {journal} {Phys. Rev. B}\ }\textbf {\bibinfo {volume} {99}},\ \bibinfo
  {pages} {054302} (\bibinfo {year} {2019})}\BibitemShut {NoStop}%
\bibitem [{\citenamefont {Jafari}(2019)}]{SciRepJafari}%
  \BibitemOpen
  \bibfield  {author} {\bibinfo {author} {\bibfnamefont {R}~\bibnamefont
  {Jafari}},\ }\bibfield  {title} {\enquote {\bibinfo {title} {Dynamical
  quantum phase transition and quasi particle excitation},}\ }\href {\doibase
  10.1038/s41598-019-39595-3} {\bibfield  {journal} {\bibinfo  {journal} {Sci.
  Rep.}\ }\textbf {\bibinfo {volume} {9}},\ \bibinfo {pages} {2871} (\bibinfo
  {year} {2019})}\BibitemShut {NoStop}%
\bibitem [{\citenamefont {Guo}\ \emph {et~al.}(2019)\citenamefont {Guo},
  \citenamefont {Yang}, \citenamefont {Zeng}, \citenamefont {Peng},
  \citenamefont {Li}, \citenamefont {Deng}, \citenamefont {Jin}, \citenamefont
  {Chen}, \citenamefont {Zheng},\ and\ \citenamefont
  {Fan}}]{DQPTexpPhysRevApplied.11.044080}%
  \BibitemOpen
  \bibfield  {author} {\bibinfo {author} {\bibfnamefont {Xue-Yi}\ \bibnamefont
  {Guo}}, \bibinfo {author} {\bibfnamefont {Chao}\ \bibnamefont {Yang}},
  \bibinfo {author} {\bibfnamefont {Yu}~\bibnamefont {Zeng}}, \bibinfo {author}
  {\bibfnamefont {Yi}~\bibnamefont {Peng}}, \bibinfo {author} {\bibfnamefont
  {He-Kang}\ \bibnamefont {Li}}, \bibinfo {author} {\bibfnamefont {Hui}\
  \bibnamefont {Deng}}, \bibinfo {author} {\bibfnamefont {Yi-Rong}\
  \bibnamefont {Jin}}, \bibinfo {author} {\bibfnamefont {Shu}\ \bibnamefont
  {Chen}}, \bibinfo {author} {\bibfnamefont {Dongning}\ \bibnamefont {Zheng}},
  \ and\ \bibinfo {author} {\bibfnamefont {Heng}\ \bibnamefont {Fan}},\
  }\bibfield  {title} {\enquote {\bibinfo {title} {Observation of a dynamical
  quantum phase transition by a superconducting qubit simulation},}\ }\href
  {\doibase 10.1103/PhysRevApplied.11.044080} {\bibfield  {journal} {\bibinfo
  {journal} {Phys. Rev. Applied}\ }\textbf {\bibinfo {volume} {11}},\ \bibinfo
  {pages} {044080} (\bibinfo {year} {2019})}\BibitemShut {NoStop}%
\bibitem [{\citenamefont {Zauner-Stauber}\ and\ \citenamefont
  {Halimeh}(2017)}]{DQPTlongrangePhysRevE.96.062118}%
  \BibitemOpen
  \bibfield  {author} {\bibinfo {author} {\bibfnamefont {Valentin}\
  \bibnamefont {Zauner-Stauber}}\ and\ \bibinfo {author} {\bibfnamefont
  {Jad~C.}\ \bibnamefont {Halimeh}},\ }\bibfield  {title} {\enquote {\bibinfo
  {title} {Probing the anomalous dynamical phase in long-range quantum spin
  chains through fisher-zero lines},}\ }\href {\doibase
  10.1103/PhysRevE.96.062118} {\bibfield  {journal} {\bibinfo  {journal} {Phys.
  Rev. E}\ }\textbf {\bibinfo {volume} {96}},\ \bibinfo {pages} {062118}
  (\bibinfo {year} {2017})}\BibitemShut {NoStop}%
\bibitem [{\citenamefont {Jurcevic}\ \emph {et~al.}(2017)\citenamefont
  {Jurcevic}, \citenamefont {Shen}, \citenamefont {Hauke}, \citenamefont
  {Maier}, \citenamefont {Brydges}, \citenamefont {Hempel}, \citenamefont
  {Lanyon}, \citenamefont {Heyl}, \citenamefont {Blatt},\ and\ \citenamefont
  {Roos}}]{heyl-trappexp}%
  \BibitemOpen
  \bibfield  {author} {\bibinfo {author} {\bibfnamefont {P.}~\bibnamefont
  {Jurcevic}}, \bibinfo {author} {\bibfnamefont {H.}~\bibnamefont {Shen}},
  \bibinfo {author} {\bibfnamefont {P.}~\bibnamefont {Hauke}}, \bibinfo
  {author} {\bibfnamefont {C.}~\bibnamefont {Maier}}, \bibinfo {author}
  {\bibfnamefont {T.}~\bibnamefont {Brydges}}, \bibinfo {author} {\bibfnamefont
  {C.}~\bibnamefont {Hempel}}, \bibinfo {author} {\bibfnamefont {B.~P.}\
  \bibnamefont {Lanyon}}, \bibinfo {author} {\bibfnamefont {M.}~\bibnamefont
  {Heyl}}, \bibinfo {author} {\bibfnamefont {R.}~\bibnamefont {Blatt}}, \ and\
  \bibinfo {author} {\bibfnamefont {C.~F.}\ \bibnamefont {Roos}},\ }\bibfield
  {title} {\enquote {\bibinfo {title} {Direct observation of dynamical quantum
  phase transitions in an interacting many-body system},}\ }\href {\doibase
  10.1103/PhysRevLett.119.080501} {\bibfield  {journal} {\bibinfo  {journal}
  {Phys. Rev. Lett.}\ }\textbf {\bibinfo {volume} {119}},\ \bibinfo {pages}
  {080501} (\bibinfo {year} {2017})}\BibitemShut {NoStop}%
\bibitem [{\citenamefont {Homrighausen}\ \emph {et~al.}(2017)\citenamefont
  {Homrighausen}, \citenamefont {Abeling}, \citenamefont {Zauner-Stauber},\
  and\ \citenamefont {Halimeh}}]{DQPTlograngisingPhysRevB.96.104436}%
  \BibitemOpen
  \bibfield  {author} {\bibinfo {author} {\bibfnamefont {Ingo}\ \bibnamefont
  {Homrighausen}}, \bibinfo {author} {\bibfnamefont {Nils~O.}\ \bibnamefont
  {Abeling}}, \bibinfo {author} {\bibfnamefont {Valentin}\ \bibnamefont
  {Zauner-Stauber}}, \ and\ \bibinfo {author} {\bibfnamefont {Jad~C.}\
  \bibnamefont {Halimeh}},\ }\bibfield  {title} {\enquote {\bibinfo {title}
  {Anomalous dynamical phase in quantum spin chains with long-range
  interactions},}\ }\href {\doibase 10.1103/PhysRevB.96.104436} {\bibfield
  {journal} {\bibinfo  {journal} {Phys. Rev. B}\ }\textbf {\bibinfo {volume}
  {96}},\ \bibinfo {pages} {104436} (\bibinfo {year} {2017})}\BibitemShut
  {NoStop}%
\bibitem [{\citenamefont {Hoyos}\ \emph
  {et~al.}(2022{\natexlab{a}})\citenamefont {Hoyos}, \citenamefont {Costa},\
  and\ \citenamefont {Xavier}}]{NetoRafaelXavierPRBL2022}%
  \BibitemOpen
  \bibfield  {author} {\bibinfo {author} {\bibfnamefont {Jos\'e~A.}\
  \bibnamefont {Hoyos}}, \bibinfo {author} {\bibfnamefont {R.~F.~P.}\
  \bibnamefont {Costa}}, \ and\ \bibinfo {author} {\bibfnamefont {J.~C.}\
  \bibnamefont {Xavier}},\ }\bibfield  {title} {\enquote {\bibinfo {title}
  {Disorder-induced dynamical griffiths singularities after certain quantum
  quenches},}\ }\href {\doibase 10.1103/PhysRevB.106.L140201} {\bibfield
  {journal} {\bibinfo  {journal} {Phys. Rev. B}\ }\textbf {\bibinfo {volume}
  {106}},\ \bibinfo {pages} {L140201} (\bibinfo {year}
  {2022}{\natexlab{a}})}\BibitemShut {NoStop}%
\bibitem [{\citenamefont {Heyl}(2018)}]{dynampt}%
  \BibitemOpen
  \bibfield  {author} {\bibinfo {author} {\bibfnamefont {Markus}\ \bibnamefont
  {Heyl}},\ }\bibfield  {title} {\enquote {\bibinfo {title} {Dynamical quantum
  phase transitions: a review},}\ }\href {\doibase 10.1088/1361-6633/aaaf9a}
  {\bibfield  {journal} {\bibinfo  {journal} {Reports on Progress in Physics}\
  }\textbf {\bibinfo {volume} {81}},\ \bibinfo {pages} {054001} (\bibinfo
  {year} {2018})}\BibitemShut {NoStop}%
\bibitem [{\citenamefont {Islam}\ \emph {et~al.}(2013)\citenamefont {Islam},
  \citenamefont {Senko}, \citenamefont {Campbell}, \citenamefont {Korenblit},
  \citenamefont {Smith}, \citenamefont {Lee}, \citenamefont {Edwards},
  \citenamefont {Wang}, \citenamefont {Freericks},\ and\ \citenamefont
  {Monroe}}]{IslamTrapIons2013}%
  \BibitemOpen
  \bibfield  {author} {\bibinfo {author} {\bibfnamefont {R.}~\bibnamefont
  {Islam}}, \bibinfo {author} {\bibfnamefont {C.}~\bibnamefont {Senko}},
  \bibinfo {author} {\bibfnamefont {W.~C.}\ \bibnamefont {Campbell}}, \bibinfo
  {author} {\bibfnamefont {S.}~\bibnamefont {Korenblit}}, \bibinfo {author}
  {\bibfnamefont {J.}~\bibnamefont {Smith}}, \bibinfo {author} {\bibfnamefont
  {A.}~\bibnamefont {Lee}}, \bibinfo {author} {\bibfnamefont {E.~E.}\
  \bibnamefont {Edwards}}, \bibinfo {author} {\bibfnamefont {C.-C.~J.}\
  \bibnamefont {Wang}}, \bibinfo {author} {\bibfnamefont {J.~K.}\ \bibnamefont
  {Freericks}}, \ and\ \bibinfo {author} {\bibfnamefont {C.}~\bibnamefont
  {Monroe}},\ }\bibfield  {title} {\enquote {\bibinfo {title} {Emergence and
  frustration of magnetism with variable-range interactions in a quantum
  simulator},}\ }\href {\doibase 10.1126/science.1232296} {\bibfield  {journal}
  {\bibinfo  {journal} {Science}\ }\textbf {\bibinfo {volume} {340}},\ \bibinfo
  {pages} {583} (\bibinfo {year} {2013})}\BibitemShut {NoStop}%
\bibitem [{\citenamefont {Porras}\ and\ \citenamefont
  {Cirac}(2004)}]{PorrasCiracTrapIons}%
  \BibitemOpen
  \bibfield  {author} {\bibinfo {author} {\bibfnamefont {D.}~\bibnamefont
  {Porras}}\ and\ \bibinfo {author} {\bibfnamefont {J.~I.}\ \bibnamefont
  {Cirac}},\ }\bibfield  {title} {\enquote {\bibinfo {title} {Effective quantum
  spin systems with trapped ions},}\ }\href {\doibase
  10.1103/PhysRevLett.92.207901} {\bibfield  {journal} {\bibinfo  {journal}
  {Phys. Rev. Lett.}\ }\textbf {\bibinfo {volume} {92}},\ \bibinfo {pages}
  {207901} (\bibinfo {year} {2004})}\BibitemShut {NoStop}%
\bibitem [{\citenamefont {Monroe}\ \emph {et~al.}(2021)\citenamefont {Monroe},
  \citenamefont {Campbell}, \citenamefont {Duan}, \citenamefont {Gong},
  \citenamefont {Gorshkov}, \citenamefont {Hess}, \citenamefont {Islam},
  \citenamefont {Kim}, \citenamefont {Linke}, \citenamefont {Pagano},
  \citenamefont {Richerme}, \citenamefont {Senko},\ and\ \citenamefont
  {Yao}}]{RMPTrap2021}%
  \BibitemOpen
  \bibfield  {author} {\bibinfo {author} {\bibfnamefont {C.}~\bibnamefont
  {Monroe}}, \bibinfo {author} {\bibfnamefont {W.~C.}\ \bibnamefont
  {Campbell}}, \bibinfo {author} {\bibfnamefont {L.-M.}\ \bibnamefont {Duan}},
  \bibinfo {author} {\bibfnamefont {Z.-X.}\ \bibnamefont {Gong}}, \bibinfo
  {author} {\bibfnamefont {A.~V.}\ \bibnamefont {Gorshkov}}, \bibinfo {author}
  {\bibfnamefont {P.~W.}\ \bibnamefont {Hess}}, \bibinfo {author}
  {\bibfnamefont {R.}~\bibnamefont {Islam}}, \bibinfo {author} {\bibfnamefont
  {K.}~\bibnamefont {Kim}}, \bibinfo {author} {\bibfnamefont {N.~M.}\
  \bibnamefont {Linke}}, \bibinfo {author} {\bibfnamefont {G.}~\bibnamefont
  {Pagano}}, \bibinfo {author} {\bibfnamefont {P.}~\bibnamefont {Richerme}},
  \bibinfo {author} {\bibfnamefont {C.}~\bibnamefont {Senko}}, \ and\ \bibinfo
  {author} {\bibfnamefont {N.~Y.}\ \bibnamefont {Yao}},\ }\bibfield  {title}
  {\enquote {\bibinfo {title} {Programmable quantum simulations of spin systems
  with trapped ions},}\ }\href {\doibase 10.1103/RevModPhys.93.025001}
  {\bibfield  {journal} {\bibinfo  {journal} {Rev. Mod. Phys.}\ }\textbf
  {\bibinfo {volume} {93}},\ \bibinfo {pages} {025001} (\bibinfo {year}
  {2021})}\BibitemShut {NoStop}%
\bibitem [{\citenamefont {Marino}\ \emph {et~al.}(2022)\citenamefont {Marino},
  \citenamefont {Eckstein}, \citenamefont {Foster},\ and\ \citenamefont
  {Rey}}]{Marino_2022Review}%
  \BibitemOpen
  \bibfield  {author} {\bibinfo {author} {\bibfnamefont {Jamir}\ \bibnamefont
  {Marino}}, \bibinfo {author} {\bibfnamefont {Martin}\ \bibnamefont
  {Eckstein}}, \bibinfo {author} {\bibfnamefont {Matthew~S}\ \bibnamefont
  {Foster}}, \ and\ \bibinfo {author} {\bibfnamefont {Ana~Maria}\ \bibnamefont
  {Rey}},\ }\bibfield  {title} {\enquote {\bibinfo {title} {Dynamical phase
  transitions in the collisionless pre-thermal states of isolated quantum
  systems: theory and experiments},}\ }\href {\doibase
  10.1088/1361-6633/ac906c} {\bibfield  {journal} {\bibinfo  {journal} {Reports
  on Progress in Physics}\ }\textbf {\bibinfo {volume} {85}},\ \bibinfo {pages}
  {116001} (\bibinfo {year} {2022})}\BibitemShut {NoStop}%
\bibitem [{\citenamefont {Zhang}\ \emph {et~al.}(2017)\citenamefont {Zhang},
  \citenamefont {Pagano}, \citenamefont {Hess}, \citenamefont {Kyprianidis},
  \citenamefont {Becker}, \citenamefont {Kaplan}, \citenamefont {Gorshkov},
  \citenamefont {Gong},\ and\ \citenamefont {Monroe}}]{ZhangTrapIons2017}%
  \BibitemOpen
  \bibfield  {author} {\bibinfo {author} {\bibfnamefont {J.}~\bibnamefont
  {Zhang}}, \bibinfo {author} {\bibfnamefont {G.}~\bibnamefont {Pagano}},
  \bibinfo {author} {\bibfnamefont {P.~W.}\ \bibnamefont {Hess}}, \bibinfo
  {author} {\bibfnamefont {A.}~\bibnamefont {Kyprianidis}}, \bibinfo {author}
  {\bibfnamefont {P.}~\bibnamefont {Becker}}, \bibinfo {author} {\bibfnamefont
  {H.}~\bibnamefont {Kaplan}}, \bibinfo {author} {\bibfnamefont {A.~V.}\
  \bibnamefont {Gorshkov}}, \bibinfo {author} {\bibfnamefont {Z.-X.}\
  \bibnamefont {Gong}}, \ and\ \bibinfo {author} {\bibfnamefont
  {C.}~\bibnamefont {Monroe}},\ }\bibfield  {title} {\enquote {\bibinfo {title}
  {Observation of a many-body dynamical phase transition with a 53-qubit
  quantum simulator},}\ }\href {\doibase 10.1038/nature24654} {\bibfield
  {journal} {\bibinfo  {journal} {Nature}\ }\textbf {\bibinfo {volume} {551}},\
  \bibinfo {pages} {551} (\bibinfo {year} {2017})}\BibitemShut {NoStop}%
\bibitem [{\citenamefont {Joshi}\ \emph {et~al.}(2022)\citenamefont {Joshi},
  \citenamefont {Schuckert}, \citenamefont {Lovas}, \citenamefont {Maier},
  \citenamefont {Blatt},\ and\ \citenamefont {Knap}}]{JoshiTrapIons2022}%
  \BibitemOpen
  \bibfield  {author} {\bibinfo {author} {\bibfnamefont {M.~K.}\ \bibnamefont
  {Joshi}}, \bibinfo {author} {\bibfnamefont {A.}~\bibnamefont {Schuckert}},
  \bibinfo {author} {\bibfnamefont {I.}~\bibnamefont {Lovas}}, \bibinfo
  {author} {\bibfnamefont {C.}~\bibnamefont {Maier}}, \bibinfo {author}
  {\bibfnamefont {R.}~\bibnamefont {Blatt}}, \ and\ \bibinfo {author}
  {\bibfnamefont {C.~F.}\ \bibnamefont {Knap}, \bibfnamefont {Roos}},\
  }\bibfield  {title} {\enquote {\bibinfo {title} {Observing emergent
  hydrodynamics in a long-range quantum magnet},}\ }\href {\doibase
  10.1126/science.abk2400} {\bibfield  {journal} {\bibinfo  {journal}
  {Science}\ }\textbf {\bibinfo {volume} {376}},\ \bibinfo {pages} {720}
  (\bibinfo {year} {2022})}\BibitemShut {NoStop}%
\bibitem [{\citenamefont {Zunkovic}\ \emph {et~al.}(2016)\citenamefont
  {Zunkovic}, \citenamefont {Silva},\ and\ \citenamefont
  {Fabrizio}}]{Bojan-LMGmodel}%
  \BibitemOpen
  \bibfield  {author} {\bibinfo {author} {\bibfnamefont {B.}~\bibnamefont
  {Zunkovic}}, \bibinfo {author} {\bibfnamefont {A.}~\bibnamefont {Silva}}, \
  and\ \bibinfo {author} {\bibfnamefont {M.}~\bibnamefont {Fabrizio}},\
  }\bibfield  {title} {\enquote {\bibinfo {title} {Dynamical phase transitions
  and loschmidt echo in the infinite-range xy model},}\ }\href {\doibase
  10.1098/rsta.2015.0160} {\bibfield  {journal} {\bibinfo  {journal} {Phil.
  Trans. R. Soc. A.}\ }\textbf {\bibinfo {volume} {374}},\ \bibinfo {pages}
  {20150160} (\bibinfo {year} {2016})}\BibitemShut {NoStop}%
\bibitem [{\citenamefont {Kosior}\ and\ \citenamefont
  {Sacha}(2018)}]{Kosier-LMGmodel}%
  \BibitemOpen
  \bibfield  {author} {\bibinfo {author} {\bibfnamefont {Arkadiusz}\
  \bibnamefont {Kosior}}\ and\ \bibinfo {author} {\bibfnamefont {Krzysztof}\
  \bibnamefont {Sacha}},\ }\bibfield  {title} {\enquote {\bibinfo {title}
  {Dynamical quantum phase transitions in discrete time crystals},}\ }\href
  {\doibase 10.1103/PhysRevA.97.053621} {\bibfield  {journal} {\bibinfo
  {journal} {Phys. Rev. A}\ }\textbf {\bibinfo {volume} {97}},\ \bibinfo
  {pages} {053621} (\bibinfo {year} {2018})}\BibitemShut {NoStop}%
\bibitem [{\citenamefont {{M. Suzuki}}(1971)}]{suzukixy}%
  \BibitemOpen
  \bibfield  {author} {\bibinfo {author} {\bibnamefont {{M. Suzuki}}},\
  }\bibfield  {title} {\enquote {\bibinfo {title} {The dimer problem and the
  generalized {X}-model},}\ }\href
  {https://doi.org/10.1016/0375-9601(71)90901-7} {\bibfield  {journal}
  {\bibinfo  {journal} {Phys. Lett. A}\ }\textbf {\bibinfo {volume} {34}},\
  \bibinfo {pages} {338} (\bibinfo {year} {1971})}\BibitemShut {NoStop}%
\bibitem [{\citenamefont {Eloy}\ and\ \citenamefont
  {Xavier}(2012)}]{Dalsonxavier}%
  \BibitemOpen
  \bibfield  {author} {\bibinfo {author} {\bibfnamefont {D.}~\bibnamefont
  {Eloy}}\ and\ \bibinfo {author} {\bibfnamefont {J.~C.}\ \bibnamefont
  {Xavier}},\ }\bibfield  {title} {\enquote {\bibinfo {title} {Entanglement
  entropy of the low-lying excited states and critical properties of an exactly
  solvable two-leg spin ladder with three-spin interactions},}\ }\href
  {\doibase 10.1103/PhysRevB.86.064421} {\bibfield  {journal} {\bibinfo
  {journal} {Phys. Rev. B}\ }\textbf {\bibinfo {volume} {86}},\ \bibinfo
  {pages} {064421} (\bibinfo {year} {2012})}\BibitemShut {NoStop}%
\bibitem [{\citenamefont {Jones}(2022)}]{LongRangeJones}%
  \BibitemOpen
  \bibfield  {author} {\bibinfo {author} {\bibfnamefont {Nick~G.}\ \bibnamefont
  {Jones}},\ }\bibfield  {title} {\enquote {\bibinfo {title} {Symmetry-resolved
  entanglement entropy in critical free-fermion chains},}\ }\href {\doibase
  10.1007/s10955-022-02941-3} {\bibfield  {journal} {\bibinfo  {journal} {J
  Stat Phys}\ }\textbf {\bibinfo {volume} {188}},\ \bibinfo {pages} {28}
  (\bibinfo {year} {2022})}\BibitemShut {NoStop}%
\bibitem [{\citenamefont {Dutta}\ and\ \citenamefont
  {Dutta}(2017)}]{KitaevLongRangQDPT2017}%
  \BibitemOpen
  \bibfield  {author} {\bibinfo {author} {\bibfnamefont {Anirban}\ \bibnamefont
  {Dutta}}\ and\ \bibinfo {author} {\bibfnamefont {Amit}\ \bibnamefont
  {Dutta}},\ }\bibfield  {title} {\enquote {\bibinfo {title} {Probing the role
  of long-range interactions in the dynamics of a long-range kitaev chain},}\
  }\href {\doibase 10.1103/PhysRevB.96.125113} {\bibfield  {journal} {\bibinfo
  {journal} {Phys. Rev. B}\ }\textbf {\bibinfo {volume} {96}},\ \bibinfo
  {pages} {125113} (\bibinfo {year} {2017})}\BibitemShut {NoStop}%
\bibitem [{\citenamefont {Defenu}\ \emph {et~al.}(2019)\citenamefont {Defenu},
  \citenamefont {Enss},\ and\ \citenamefont
  {Halimeh}}]{KitaevLongRangeHalimeh}%
  \BibitemOpen
  \bibfield  {author} {\bibinfo {author} {\bibfnamefont {Nicol\`o}\
  \bibnamefont {Defenu}}, \bibinfo {author} {\bibfnamefont {Tilman}\
  \bibnamefont {Enss}}, \ and\ \bibinfo {author} {\bibfnamefont {Jad~C.}\
  \bibnamefont {Halimeh}},\ }\bibfield  {title} {\enquote {\bibinfo {title}
  {Dynamical criticality and domain-wall coupling in long-range
  hamiltonians},}\ }\href {\doibase 10.1103/PhysRevB.100.014434} {\bibfield
  {journal} {\bibinfo  {journal} {Phys. Rev. B}\ }\textbf {\bibinfo {volume}
  {100}},\ \bibinfo {pages} {014434} (\bibinfo {year} {2019})}\BibitemShut
  {NoStop}%
\bibitem [{\citenamefont {Uhrich}\ \emph {et~al.}(2020)\citenamefont {Uhrich},
  \citenamefont {Defenu}, \citenamefont {Jafari},\ and\ \citenamefont
  {Halimeh}}]{KitaevLongHalimedsupercon}%
  \BibitemOpen
  \bibfield  {author} {\bibinfo {author} {\bibfnamefont {Philipp}\ \bibnamefont
  {Uhrich}}, \bibinfo {author} {\bibfnamefont {Nicol\`o}\ \bibnamefont
  {Defenu}}, \bibinfo {author} {\bibfnamefont {Rouhollah}\ \bibnamefont
  {Jafari}}, \ and\ \bibinfo {author} {\bibfnamefont {Jad~C.}\ \bibnamefont
  {Halimeh}},\ }\bibfield  {title} {\enquote {\bibinfo {title}
  {Out-of-equilibrium phase diagram of long-range superconductors},}\ }\href
  {\doibase 10.1103/PhysRevB.101.245148} {\bibfield  {journal} {\bibinfo
  {journal} {Phys. Rev. B}\ }\textbf {\bibinfo {volume} {101}},\ \bibinfo
  {pages} {245148} (\bibinfo {year} {2020})}\BibitemShut {NoStop}%
\bibitem [{\citenamefont {Su}\ \emph {et~al.}(1979)\citenamefont {Su},
  \citenamefont {Schrieffer},\ and\ \citenamefont {Heeger}}]{SShmodel}%
  \BibitemOpen
  \bibfield  {author} {\bibinfo {author} {\bibfnamefont {W.~P.}\ \bibnamefont
  {Su}}, \bibinfo {author} {\bibfnamefont {J.~R.}\ \bibnamefont {Schrieffer}},
  \ and\ \bibinfo {author} {\bibfnamefont {A.~J.}\ \bibnamefont {Heeger}},\
  }\bibfield  {title} {\enquote {\bibinfo {title} {Solitons in
  polyacetylene},}\ }\href {\doibase 10.1103/PhysRevLett.42.1698} {\bibfield
  {journal} {\bibinfo  {journal} {Phys. Rev. Lett.}\ }\textbf {\bibinfo
  {volume} {42}},\ \bibinfo {pages} {1698--1701} (\bibinfo {year}
  {1979})}\BibitemShut {NoStop}%
\bibitem [{\citenamefont {Ares}\ \emph {et~al.}(2022)\citenamefont {Ares},
  \citenamefont {Murciano},\ and\ \citenamefont
  {Calabrese}}]{Ares_Calabrese2022LongRange}%
  \BibitemOpen
  \bibfield  {author} {\bibinfo {author} {\bibfnamefont {Filiberto}\
  \bibnamefont {Ares}}, \bibinfo {author} {\bibfnamefont {Sara}\ \bibnamefont
  {Murciano}}, \ and\ \bibinfo {author} {\bibfnamefont {Pasquale}\ \bibnamefont
  {Calabrese}},\ }\bibfield  {title} {\enquote {\bibinfo {title}
  {Symmetry-resolved entanglement in a long-range free-fermion chain},}\ }\href
  {\doibase 10.1088/1742-5468/ac7644} {\bibfield  {journal} {\bibinfo
  {journal} {Journal of Statistical Mechanics: Theory and Experiment}\ }\textbf
  {\bibinfo {volume} {2022}},\ \bibinfo {pages} {063104} (\bibinfo {year}
  {2022})}\BibitemShut {NoStop}%
\bibitem [{\citenamefont {Hoyos}\ \emph
  {et~al.}(2022{\natexlab{b}})\citenamefont {Hoyos}, \citenamefont {Xavier},\
  and\ \citenamefont {Costa}}]{NetoRafaelXavierunpublish}%
  \BibitemOpen
  \bibfield  {author} {\bibinfo {author} {\bibfnamefont {Jos\'e~A.}\
  \bibnamefont {Hoyos}}, \bibinfo {author} {\bibfnamefont {J.~C.}\ \bibnamefont
  {Xavier}}, \ and\ \bibinfo {author} {\bibfnamefont {R.~F.~P.}\ \bibnamefont
  {Costa}},\ }\bibfield  {title} {\enquote {\bibinfo {title} {
  Dynamical Griffiths singularities in certain random spin chains},}\
  }\href@noop {} {\bibfield  {journal} {\bibinfo  {journal} {(unpublished)}\ }
  }\BibitemShut {NoStop}%
\bibitem [{\citenamefont {Jaeger}(1998)}]{Ehrenfest}%
  \BibitemOpen
  \bibfield  {author} {\bibinfo {author} {\bibfnamefont {Gregg}\ \bibnamefont
  {Jaeger}},\ }\bibfield  {title} {\enquote {\bibinfo {title} {Classification
  of phase transitions: Introduction and evolution},}\ }\href {\doibase
  10.1007/s004070050021} {\bibfield  {journal} {\bibinfo  {journal} {Arch Hist
  Exact Sc.}\ }\textbf {\bibinfo {volume} {53}},\ \bibinfo {pages} {51--81}
  (\bibinfo {year} {1998})}\BibitemShut {NoStop}%
\bibitem [{\citenamefont {Schmitt}\ and\ \citenamefont
  {Kehrein}(2015)}]{schmitt-kehrein-prb15}%
  \BibitemOpen
  \bibfield  {author} {\bibinfo {author} {\bibfnamefont {M.}~\bibnamefont
  {Schmitt}}\ and\ \bibinfo {author} {\bibfnamefont {S.}~\bibnamefont
  {Kehrein}},\ }\bibfield  {title} {\enquote {\bibinfo {title} {Dynamical
  quantum phase transitions in the kitaev honeycomb model},}\ }\href {\doibase
  10.1103/PhysRevB.92.075114} {\bibfield  {journal} {\bibinfo  {journal} {Phys.
  Rev. B}\ }\textbf {\bibinfo {volume} {92}},\ \bibinfo {pages} {075114}
  (\bibinfo {year} {2015})}\BibitemShut {NoStop}%
\bibitem [{\citenamefont {Byers}\ and\ \citenamefont
  {Yang}(1961)}]{Twist1-PhysRevLett.7.46}%
  \BibitemOpen
  \bibfield  {author} {\bibinfo {author} {\bibfnamefont {N.}~\bibnamefont
  {Byers}}\ and\ \bibinfo {author} {\bibfnamefont {C.~N.}\ \bibnamefont
  {Yang}},\ }\bibfield  {title} {\enquote {\bibinfo {title} {Theoretical
  considerations concerning quantized magnetic flux in superconducting
  cylinders},}\ }\href {\doibase 10.1103/PhysRevLett.7.46} {\bibfield
  {journal} {\bibinfo  {journal} {Phys. Rev. Lett.}\ }\textbf {\bibinfo
  {volume} {7}},\ \bibinfo {pages} {46--49} (\bibinfo {year}
  {1961})}\BibitemShut {NoStop}%
\bibitem [{\citenamefont {Kohn}(1964)}]{Twist2-PhysRev.133.A171}%
  \BibitemOpen
  \bibfield  {author} {\bibinfo {author} {\bibfnamefont {Walter}\ \bibnamefont
  {Kohn}},\ }\bibfield  {title} {\enquote {\bibinfo {title} {Theory of the
  insulating state},}\ }\href {\doibase 10.1103/PhysRev.133.A171} {\bibfield
  {journal} {\bibinfo  {journal} {Phys. Rev.}\ }\textbf {\bibinfo {volume}
  {133}},\ \bibinfo {pages} {A171--A181} (\bibinfo {year} {1964})}\BibitemShut
  {NoStop}%
\bibitem [{\citenamefont {Poilblanc}(1991)}]{twist3-PhysRevB.44.9562}%
  \BibitemOpen
  \bibfield  {author} {\bibinfo {author} {\bibfnamefont {Didier}\ \bibnamefont
  {Poilblanc}},\ }\bibfield  {title} {\enquote {\bibinfo {title} {Twisted
  boundary conditions in cluster calculations of the optical conductivity in
  two-dimensional lattice models},}\ }\href {\doibase 10.1103/PhysRevB.44.9562}
  {\bibfield  {journal} {\bibinfo  {journal} {Phys. Rev. B}\ }\textbf {\bibinfo
  {volume} {44}},\ \bibinfo {pages} {9562--9581} (\bibinfo {year}
  {1991})}\BibitemShut {NoStop}%
\end{thebibliography}%

\end{document}